\documentclass[aps,superscriptaddress,showpacs,nofootinbib]{revtex4}
\usepackage[dvips]{graphics}
\usepackage[dvips]{graphicx}
\usepackage{bm}

\input{epsf}

\newcommand{\be}{\begin{equation}}
\newcommand{\ee}{\end{equation}}
\newcommand{\bea}{\begin{eqnarray}}
\newcommand{\eea}{\end{eqnarray}}

\begin{document}

\preprint{UGFT-218/07}

\preprint{CAFPE-88/07}

\title{Non-linear Preheating with Scalar Metric Perturbations}
\author{Mar Bastero-Gil}
\email{mbg@ugr.es}
\affiliation{Departamento de F\'{\i}sica Te\'orica y del Cosmos,
  Universidad de Granada, Granada-18071, Spain}

\author{M. Tristram}
\email{tristram@lal.in2p3.fr}
\affiliation{ LPSC, Universit\'e Joseph Fourier Grenoble 1,
  CNRS/IN2P3, Institut Polytechnique de Grenoble, Grenoble, France
}
\affiliation{ LAL, Univ Paris-Sud, CNRS/IN2P3, Orsay, France }

\author{J. Macias-P\'erez}
\email{macias@lpsc.in2p3.fr}
\affiliation{ LPSC, Universit\'e Joseph Fourier Grenoble 1,
  CNRS/IN2P3, Institut Polytechnique de Grenoble, Grenoble, France
}

\author{D. Santos}
\email{santos@lpsc.in2p3.fr}
\affiliation{ LPSC, Universit\'e Joseph Fourier Grenoble 1,
  CNRS/IN2P3, Institut Polytechnique de Grenoble, Grenoble, France
}

\begin{abstract}

We have studied preheating of field perturbations in a 3-dimensional lattice
including the effect of scalar metric perturbations, in two generic 
models of inflation: chaotic inflation with a quartic potential, and
standard hybrid inflation. We have prepared the initial state for the classical
evolution of the system with vanishing vector and tensor metric
perturbations, consistent with the constraint equations, the
energy and momentum constraints. The non-linear evolution 
inevitably generates vector and tensor modes, and this reflects on how
well the constraint equations are fulfilled during the evolution. 
The induced preheating of the scalar metric
perturbations is not large enough to backreact onto the fields, but it
could affect the evolution of vector and tensor modes. 
This is the case in hybrid inflation for some values of the coupling
$g$  and the height of potential $V_0^{1/4}$. For example with 
$V_0^{1/4} \simeq  10^{15}$ GeV, preheating of scalar perturbations is
such that their source term in the evolution equation for tensor and
vector fluctuations becomes comparable to that of the field
anisotropic stress.

\medskip                                   

\noindent
keywords: cosmology, inflation
\end{abstract}

\maketitle

\section{Introduction}
\label{sect1}
Cosmological observations, and in particular Cosmic Microwave
Background (CMB) measurements \cite{COBE,CMB1,CMB2,CMB3,WMAP},  are consistent
with an early  period of inflation, 
which among other things accounts for the inferred flatness of the
Universe, and gives rise to the primordial curvature perturbation which
would seed the large scale structure observed today. The value of the
primordial power spectrum at large scales was first measured by COBE experiment
\cite{COBE}, and it has been confirmed by all subsequent 
CMB experiments. The  spectrum is consistent 
with a gaussian, and practically scale-invariant spectrum, although
the latest data from the Wilkinson Microwave Anisotropy Probe (WMAP)
\cite{WMAP} seems to prefer a  red-tilted spectrum when no tensor and no other
contributions are included, showing  a positive correlation
between the tensor-to-scalar ratio and the spectral index
\cite{tensor-strings}. Nevertheless, 
present data sets at most an upper limit on the level of the tensor
contribution and non-gaussianity, which even if small, could still be
observed by the next generation of CMB experiments, like ESA's Planck
surveyor satellite \cite{planck}.    
   
From the theoretical side, the quest now is for a realistic particle
physics model of inflation, and a better understanding of the
inflationary and post-inflationary dynamics. 
In order to allow for the conversion of the vacuum energy into
radiation at the end of inflation, the inflaton should couple to other
fields. In the standard picture, this process called reheating
takes place through the 
perturbative decay of the inflaton into light degrees of freedom,
which thermalize into radiation. Nevertheless, previous to the
perturbative decay, the evolution of the system may be dominated by
non-perturbative effects as those of preheating \cite{preheating,preheating2},
i.e., parametric amplification of quantum field fluctuations in a
background of oscillating fields. Through parametric resonance, field
mode amplitudes grow exponentially within certain
resonance bands in $k$ space, being this a more efficient and faster way of
transferring vacuum energy into radiation than the standard reheating
mechanism. This process and its consequences for the subsequent 
cosmological evolution has been extensively studied in the literature
for different kind of models \cite{prechaotic4, prehybrid1, prehybrid2,preothers}. The first
stages of preheating can be studied within linear perturbation theory at first
order \cite{preheating}, but soon after the resonance develops one
would need to improve on the perturbative expansion in order to take
into account backreaction and 
rescattering effects. Backreaction effects can be partially
incorporated by using the Hartree-Fock approximation  
\cite{hartree1,hartree2}. However, in order to take fully into account
rescattering effects, i.e., mode-to-mode couplings, one  has to resort to
non-perturbative tools, like lattice calculations \cite{lattice4-1,lattice4-2,latthybrid,latthybrid-2,lattice}.  
Preheating, although a very fast process
compare with the cosmological scales, lasting at most only a
few e-folds of expansion, may lead to a very rich phenomenology having
to do, among others,  with production of massive relics
\cite{prerelics}, baryogenesis and leptogenesis \cite{prebaryon,
  latthybbaryon}, and black holes production  \cite{prebh,lattice}. 

In addition, in a cosmological framework metric 
perturbations might also be parametrically amplified during 
preheating. This has been studied for example using linear
perturbation theory \cite{linearmetric, linearmetric2}, up to second order 
\cite{secondmetric,prenongauss},  and with the Hartree-Fock
approximation \cite{hartreemetric}. A relevant
question is whether or not this non-adiabatic 
excitation of field fluctuations leads to the amplification of 
super-Hubble curvature fluctuations \cite{kodama2,basset,counterexample}. 
This may happen in the presence of an entropy/isocurvature perturbation
mode, not suppressed on large scales during inflation, which sources
the curvature perturbation when parametrically amplified during
preheating \cite{preentropy}. Other approach for the study of
non-linear super-Hubble cosmological perturbations during inflation 
relies on a gradient expansion, by assuming that the spatial
derivatives are small compared to the time derivative
\cite{gradient1,gradient2}. And more recently it has been  developed a
covariant formalism which would 
allow to study the full non-linear evolution of cosmological
perturbations. It was also shown that the non-linear
covariant generalization stays constant on super-Hubble scales in the
absence of a non-adiabatic source \cite{vernizzi}. 
  
Besides the amplification of the scalar perturbations, the
non-linear nature of the preheating process can also induce some
level of primordial non-gaussianity in the spectrum, to the level
detectable by the Planck mission \cite{nongauss,prenongauss}. In 
addition, it could enhance the tensor perturbations, giving rise to a
stochastic background of gravitational waves within the reach of the
future planned gravitational observatories \cite{pregw,pregw2}. 

Therefore, preheating after inflation  can potentially alter the
inflationary predictions. In many cases, like searching
for non-gaussianity or gravitational waves, we are after very small effects
resonantly enhanced during preheating. These are by default at least
second order in perturbation 
theory, for which the mixing of the different kinds of perturbations,
scalar, vector and tensor, is unavoidable.  
For example, beyond linear perturbation, tensors are seeded not only by
scalars but also by vectors \cite{mollerach,secondordergw}, the latter
in turn being generated by the 
scalars. Aside from preheating effects, it has been shown \cite{mollerach}
that secondary vector and tensor modes could 
induced a non-vanishing B-mode polarization in the CMB. And this effect
becomes comparable or even dominates over that of primary
gravitational waves for a primordial tensor-to-scalar ratio $r \leq
10^{-6}$. This is for example the predicted ratio for models with an
inflationary energy scale of the order of  $O(10^{15})$ GeV, i.e.,
one order of magnitude below the typical grand unification scale
\cite{lythriotto}. Given 
the relevance, for inflationary model building, of the detection of a
background of gravitational waves, it is important therefore to take
into account 
all possible generating and/or amplifying mechanisms of the B-mode
polarization during cosmological evolution. In particular, it would be
interesting 
to include in the non-linear simulations of preheating the effect of
scalar, vector  and tensor metric perturbations all together. Although the
backreaction effect on the fields might be quite small, as expected,
they could for example backreact onto each other, affecting the final
tensor spectrum.     

In most of the previous studies of preheating, fields are evolved in a
background metric, i.e., neglecting the effect of metric 
perturbations themselves on the evolution of the fields. It is argued that
the backreaction effect of these perturbations on the field evolution
is small, and the approximation should hold to a good extent when
dealing with scalar perturbations. Other studies take into account
metric variables in a one dimensional system \cite{precurv1dim},  or 
reducing the system to a one dimensional system by the use 
of symmetries, either planar \cite{precurvnum1} or radial
\cite{precurvrad}. These one-dimensional studies also focus on scalar
metric fluctuations, without vectors or tensors, integrating the
system in a lattice with 
$N$ sites. 
 While using lattice techniques allows for a full non-linear treatment
 of the problem, the procedure has its own
limitations when dealing with an expanding universe. By discretizing
the space and putting the system in a box of length $L$ one
introduces both a comoving ultraviolet and infrared cut-off in the system.  
Due to the expansion of the Universe, the physical momentum is redshifted and  
the resonance bands during preheating move towards higher values of
the comoving  wavenumber. Therefore,  in order to keep the comoving ultraviolet
cut-off larger than the effective cut-off for preheating we need a
large ratio $N/L$. On the other hand, if we  
were interested in super-Hubble perturbations, the infrared cut-off
would have to be smaller than the Hubble rate of expansion at least at
the beginning of the evolution, which implies taking $L$
as large as possible. 
However, limitations on computer memory and CPU
resources prevend one from  working with too big lattices. 
The advantage of effectively reducing the
problem to a one-dimensional system lies on the possibility of
considering large enough lattices, and therefore super-hubble modes, with
less computing time cost.     
We have extended these effectively
one-dimensional studies to a 3-dimensional system, without using any
explicit symmetry. We are going to consider  
preheating of inflaton fluctuations  including
scalar metric perturbations in a cubic 3-dimensional spatial lattice with
$N^3$ sites, and length $L$. By considering  $L$ large enough 
would allow to include super-Hubble modes,  but at the expense of
quickly loosing the resonance bands during the evolution;
unless we increment $N$ accordingly, which is not viable in our case. Therefore,  as a
first step we focus on the possible 
effect of sub-Hubble metric perturbations during preheating. 

In writing down the evolution equations for the scalar field in
general relativity, we need 
first to choose a coordinate space-time $(t, {\bf x})$. The choice of
coordinates defines a threading of space-time 
(fixed ${\bf x}$) and a slicing into hypersurfaces (fixed $t$). In the
ADM formalism \cite{adm}, the spacetime line element is given by: 
\bea
ds^2 &=& g_{\mu \nu} dx^\mu d x^\nu= N^2 dt^2 -\gamma_{ij} (dx^i+ N^i dt) (dx^j+ N^j dt) \,,
\label{metric}
\eea
with $N$ the lapse function, $N^i$ the shift vector
field,  and $\gamma_{ij}$ the spatial metric. A choice of $N$ and $N^i$
fixes the gauge. Einstein equations plus the conservation of the
stress energy tensor can be split  into a set of four constraint
equations, the Hamiltonian and momentum constraints, and a set of
evolution equations  for the spatial metric and matter fields. 
The initial conditions for matter and metric variables must satisfy
the constraint equations, and cannot be freely specified. This is the well
known  initial-value problem in general relativity
\cite{initialvalue}.  Once the constraint equations are imposed
on the initial slide at $t=0$, they are fulfilled at any time $t$ by
the evolved quantities\footnote{Although the discretization procedure will
invariably introduce deviations from the constraints during the
evolution \cite{choptuik}.}. We will work in the synchronous
gauge, which means null shift vector $N^i=0$ and lapse
function $N=1$. As a first step, we will only follow the evolution of
scalar metric  perturbations, neglecting vector and tensor modes. That
is, we will choose initial conditions to ensure that the initial
vectors vanishes and the tensors are negligible, and later check to
which extent they are generated by the non-linear evolution, i.e., to
which extent the constraint equations are fulfilled.   

On the matter side, we have studied  two different and characteristic
types of inflationary models with scalar fields minimally coupled to
gravity: chaotic models with a quartic potential 
$\lambda \Phi^4$, and standard supersymmetric hybrid model. Preheating
in both of 
them has been extensively studied in the literature, including lattice
simulations. For the chaotic quartic model the resonance starts in a
single, narrow resonance band \cite{prechaotic4} at a fixed value of 
the comoving wavenumber, but later on due to rescattering  the resonance
spreads to several bands in momentum space, until finally the spectrum
is smoothed out \cite{lattice4-1}.  In hybrid models due to the
tachyonic instability at the end of inflation \cite{prehybrid2} the
lower momentum modes are quickly amplified, and in a few
oscillations of the background fields the whole process ends
\cite{latthybrid, latthybrid-2}. 

In section II we give the set of equations for metric and field
variables to be integrated in the lattice, and discuss the issue of
the initial conditions. In section III we present the results for the
chaotic model with a quartic potential;  results for the hybrid model
are given in section IV. In both cases, we compare the evolution with
and without scalar metric perturbations.  
The summary of our results is given in section V.

\section{Evolution equations and initial conditions}
Our choice of gauge reduces the metric given in Eq. (\ref{metric}) to
the line element:  
\bea
ds^2 &=& e^{2 \alpha(t,x)} dt^2 -e^{2\beta(t,x)}\tilde \gamma_{ij} dx^i dx^j \,,
\label{metric2}
\eea
where we have defined the lapse function $N=e^\alpha$, and the spatial metric 
$\gamma_{ij}=e^{2 \beta} \tilde \gamma_{ij}$, with $Det(\tilde
\gamma)=1$.  Although we are going to work in the synchronous gauge
and later set $N=e^\alpha=1$, we will keep the dependence on the lapse
function explicit in this section for the sake of generality. 
The matter content is given in general by a set of scalar fields
$\Phi_I(t,x)$, with the Lagrangian and stress-energy tensor:
\bea
{\cal L} &=& \frac{1}{2}\nabla_\mu \Phi^I \nabla^\mu \Phi_I -V(\Phi_I) \,,\\
T_{\mu \nu} &=& \nabla_\mu \Phi^I\nabla_\nu \Phi_I - \frac{1}{2} g_{\mu \nu} \left( \nabla_\sigma \Phi^I \nabla^\sigma \Phi_I + 2 V(\Phi_I) \right) \,,
\eea
where the subindex ``I'' counts the no. of fields and summation is
understood over repeated index;  $V(\Phi_I)$ is the potential for the
fields considered, and 
$\nabla_\mu$  the covariant derivative for the metric $g_{\mu \nu}$.    
With this choice, the evolution equation for each scalar is given by:
\be
\Phi_I^{\prime \prime} + 3 \beta^\prime \Phi_I^\prime - D_i D^i \Phi_I
+ V_I - D_i \alpha D^i \Phi_I =0 \,, \label{evolfield}
\ee
where now $D_i$ is the spatial covariant derivative given by the
spatial metric $\gamma_{ij}$, and we have defined:
\be
\Phi_I^{\prime}= e^{-\alpha} \partial_t \Phi_I \,\;\;\;\;
\beta^{\prime}= e^{-\alpha} \partial_t \beta \,.
\ee
In the covariant formalism \cite{covariant}, these are the
covariant time derivative along the unit four velocity $u^\mu =
dx^\mu/ds=(e^{-\alpha},0,0,0)$, with $f^\prime= u^\mu \nabla_\mu f$
for any scalar quantity. The volume expansion is then given by
$\nabla_\mu u^\mu = 3 \beta^\prime$, and the acceleration vector by
$a^\mu = u^\nu \nabla_\nu u^\mu= (0, D^i \alpha)$. Given the four
vector $u^\mu$, the stress-energy tensor can be decomposed as:
\be
T_{\mu \nu}= ( \rho +P) u_\mu u_\nu + g_{\mu \nu} P + q_\mu u_\nu
+ q_\nu u_\mu + \Pi_{\mu \nu} \,,
\ee
where $\rho$, $P$, $q_\mu$ and $\pi_{\mu \nu}$ are respectively the
energy density, 
pressure, momentum and anisotropic stress in the frame defined by
$u^\mu$, and given by:
\bea
\rho&= &\frac{1}{2} \left( \Phi_I^\prime \Phi^{I\prime} + D^i \Phi_I D_i
\Phi_I \right) + V(\phi_I) \,, \label{rho}\\ 
P&=& \frac{1}{2} \left( \Phi_I^\prime \Phi^{I\prime} - \frac{1}{3} D^i
\Phi_I D_i \Phi_I \right) - V(\phi_I) \,, \\ 
q_i&=& - \Phi_I^\prime D_i\Phi^I \,, \\
\Pi_{ij}&=& D_i \Phi^I D_j \Phi_I - \frac{1}{3} \gamma_{ij} D_k \Phi^I
D^k \Phi_I \,. 
\eea
Einstein equations give the evolution of the spatial
metric components, plus 2 constraint equations. With our choice of gauge, the
evolution equation for $\beta$ reduces to:
\be
3 \beta^{\prime \prime} + 3 (\beta^\prime)^2 - D_i D^i \alpha =- \frac{\kappa}{2} \left(\rho +3 P\right)-  A^i_j A^k_i\,, \label{evoltrace}
\ee
 where $\kappa= 8 \pi G_N= 1/m_P^2$, with 
$m_P=2.42\times 10^{18}$ GeV the reduced Planck mass, and $A^i_j$ is
 the traceless part of the extrinsic curvature $K^i_j$ \cite{adm}, which in our
 case is given by  
\bea
K^i_j &=&\frac{e^{-\alpha}}{2}\gamma^{ik}\partial_t \gamma_{kj}=
\beta^\prime \delta_j^i + A_j^i
\,,
\eea
where $A^i_j=\tilde \gamma^{ik} \tilde \gamma^\prime_{kj}/2$. The
latter contains the vector, tensors and the traceless scalar mode,
while $\beta^\prime$ gives the local expansion rate. 
The equation for $A^i_j$ reads:
\be
A^{i\prime}_j + 3 \beta^\prime A^i_j= e^{-\alpha}\left( D^iD_j e^\alpha
- \frac{\delta^i_j}{3} D^kD_k e^\alpha \right) 
- \left( ^{(3)}\!R^i_{\,j} 
- \frac{\delta^i_j}{3}\,{^{(3)}\!R} \right)
+ \kappa \left( \Pi^i_j - \frac{\delta^i_j}{3} \Pi \right)  \,, \label{evoltraceless}
\ee
where $^{(3)}\!R^i_{\,j}$, $^{(3)}\!R$ are the Ricci tensor and scalar
with respect to the spatial metric. For example for a flat spatial
metric with an inhomogeneous scale factor, $\bar \gamma_{ij}= e^{2 \beta}
\delta_{ij}$, we have:
\be
^{(3)}\!\bar R^i_{\,j} 
- \frac{\delta^i_j}{3}\, ^{(3)}\!\bar R = e^{-2 \beta} \left ( \partial^i
\beta \partial_j \beta - \partial^i \partial_j \beta -
\frac{\delta^i_j}{3} ( \partial^k \beta \partial_k \beta - \partial^k
\partial_k \beta) \right) \,,
\ee
while the field source term is given by:
\be
\Pi^i_j - \frac{\delta^i_j}{3} \Pi = e^{-2 \beta}\left(\partial^i
\Phi \partial_j \Phi - \frac{\delta^i_j}{3} \partial^k \Phi \partial_k
\Phi \right) \,.
\ee
The Hamiltonian and the momentum constraint are  
given respectively by: 
\bea
6 \beta^{\prime^2} - ^{(3)}\!\!R - A^i_j A^j_i&=& 2\kappa \rho \,,
\label{hubble}\\ 
2 D_i \beta^\prime - D_j A^j_i &=& -\kappa\Phi_I^\prime D_i \Phi^I
\,.\label{momentum} 
\eea

The full set of equations (\ref{evolfield}), (\ref{evoltrace}) and
(\ref{evoltraceless}) have been integrated in a lattice in
Ref. \cite{Kurki} for a chaotic
inflation model, where it was shown the viability of inflation
starting with inhomogeneous initial conditions. We will instead
follow the evolution starting just at the end of inflation. However,
due to the complexity of the problem, as a first step we will only keep
the effect of the scalar metric variable $\beta$, assuming that
vector and tensors are subdominant and negligible. Afterward we will
cross-check the  
consistency of this approximation. Therefore, keeping only the $\beta$
dependent terms the final set of equations to be integrated is given
by: 
\bea
\Phi_I^{\prime \prime} + 3 \beta^\prime \Phi_I^\prime - e^{-2 \beta
}\partial^i \partial _i \Phi_I
+ V_I - e^{-2 \beta }( \partial^i \alpha + \partial^i \beta)
\partial_i \Phi_I &\simeq& 0 \,, \label{evolfield2} \\
 3 \beta^{\prime \prime} + 3 (\beta^\prime)^2 - e^{-2 \beta }(\partial
 ^i \partial_i \alpha + \partial
 ^i \alpha \partial_i \alpha + \partial
 ^i \beta \partial_i \alpha) &\simeq&  -\frac{\kappa}{2} \left(\rho +3 P\right)
 \,. \label{evoltrace2} 
\eea

The initial conditions for $\Phi$, $\Phi^\prime$,
$\beta$, $\beta^\prime$ must be compatible with the constraints. 
For the fields we follow the standard
procedure, with the quantum field theory being replaced by an
equivalent classical field theory. Equivalent in the sense that
the expectation values of quantum variables are equal to the
average values of the classical ones. The classical to quantum transition
takes place during inflation, when interactions of the quantum fields
can be neglected. Therefore, the field initial conditions
for the subsequent classical evolution can be obtained from the linear
theory of quantum-to-classical transition \cite{starobinsky}. 

The field is expanded in Fourier modes in a spatial lattice of volume
$L^3$ with periodic boundary conditions. The zero mode
corresponds to the 
homogeneous value of the field given by its average value in the
lattice, $\Phi(t)= \langle \Phi(t,x) \rangle$. The non-zero modes are
expanded in $k$-space as:
\bea
\Phi_I(t,x) &=& \frac{1}{V} \sum_k q_I(t,k) e^{i k x} \,, \label{phik}\\
\partial_t \Phi_I(t,x) &=& \frac{1}{V} \sum_k \partial_t q_I(t,k) e^{i k
  x} \label{phidotk}\,. 
\eea
The initial vacuum state $q(0,k)$ corresponds to a complex Gaussian
distribution with a random phase for each mode $k$,  and root mean squared
\be
|q(0,k)|_{rms}\simeq \frac{1}{\sqrt{2 \omega_k}} \,, \label{initspectrum}
\ee
 where $\omega_k= \sqrt{k^2 +
  V_{II}}$ is the initial frequency of each k-mode. The initial time
derivative for the field is then given by $\partial_t q(0,k)= -i
\omega_k(0) q(0,k)$.   

Having set the initial field's values, we have to deal now with the
initial conditions for the metric variable, such that the constraint
equations are fulfilled at $t=0$. Those are given by:
\bea
3 \beta^{\prime^2} - e^{-2 \beta} ( \partial^k \beta \partial_k \beta
+ 2 \partial^k \partial_k \beta) &=& \kappa \rho \,,
\label{hubble2}\\ 
2 \partial_i \beta^\prime &=& -\kappa\Phi_I^\prime \partial_i \Phi^I
\,. \label{momentum2}
\eea
The main restriction comes from the momentum
constraint: without including metric vector variables, the RHS of
Eq. (\ref{momentum2}) must reduce to a gradient term at least at
$t=0$. However, this is
not the case in the synchronous gauge when studying single field
models with initial conditions for  
the field  as given in Eqs. (\ref{phik}) and (\ref{phidotk}). In order
to be consistent with our assumption of negligible vector and tensor
modes, we can take 
instead an homogeneous initial profile for the field velocity, given by
its background value, i.e, $\Phi^\prime= \langle \partial_t \Phi(0,x)
\rangle$. This will set our choice of initial conditions for the
chaotic model studied in section III.  
Another option often used in the literature could be to add an extra,
non-interacting scalar field \cite{precurvnum1}. Its initial velocity
can be adjusted such that the momentum constraint is trivially satisfied
\cite{extrafield}.  The energy  density of the non-interacting extra
field would behave as radiation due to the gradient contribution in
Eq. (\ref{rho}), and setting it to be negligible at $t=0$, it would
remain so during the evolution. We will use instead this choice for
the hybrid model in section IV, where the role of the ``extra'' field
is played by the waterfall field already present in the model.   

Having adjusted the  initial field velocities in order to
fulfill the momentum constraint, Eqs. (\ref{hubble2}) and
(\ref{momentum2}) provide the initial values of $\beta$ and $\beta^\prime$ at
$t=0$. For one-dimensional systems (or effectively one-dimensional)
this procedure is consistent with neglecting vector and tensor
perturbations, and the constraints are preserved by the
evolution. However, in general this is not the case in 3 dimensions without
additional symmetries, and vectors and tensors will be induced during
the non-linear evolution. Therefore, the departure from zero of the momentum
constraint during the evolution may give us an estimation of the
amplitude for example of the vector perturbations generated during
preheating\footnote{Tensor perturbations, being transverse and
  traceless, do not contribute to the momentum constraint.}.  
On the other hand, the scalar perturbations we consider will
contribute to the source term  
for tensor perturbations, i.e, the RHS of Eq. (\ref{evoltraceless}),
and we can compare this to the pure anisotropic scalar field source term,
as those considered in \cite{pregw2}, 
and check whether or not  they may become comparable during preheating.

\section{Chaotic model}
We first study the chaotic inflation model with potential $V=\lambda
\Phi^4/4$. In this model, inflation takes place for values of the
background field larger than $m_P$, and ends approximately when $
\phi(t)=\langle \Phi(t) \rangle  \simeq 2
\sqrt{3} m_P$. Following Ref. \cite{lattice4-1}, we start evolving the
field when oscillations begin, at a slightly smaller value
$\phi_0\equiv \langle \Phi(0)\rangle = \sqrt{3} m_P$. Approximately this is the
value at which the conformal time derivative of the field vanishes. We
rescale the field by its 
initial value $\phi_0$, and time will be given in units of
$1/\sqrt{\lambda} \phi_0$. In program units, the initial background
field velocity is $\dot \phi_0  \equiv \langle \dot \Phi(0) \rangle  =
-1/\sqrt{2}$.  The comoving wavenumber is given  in
units of $\sqrt{\lambda} \phi_0$. With this rescaling the coupling
$\lambda$ does not appear in the equations of motion, but it rescales
the initial value of the fluctuations (the initial spectrum). Given
that the development of the resonance for the inflaton 
fluctuations  is not very sensitive to its value, we will take $\lambda =
10^{-4}$ for convenience, although in a realistic model it should be
$\lambda\simeq O(10^{-14})$ in order to match the amplitude of the
primordial spectrum with the observed value. 

The initial values for the field fluctuations are given by
 Eqs. (\ref{phik}), and 
 (\ref{initspectrum}), and we adjust the initial value of the field
 velocity  in order to fulfill the momentum constraint at $t=0$, such that
\be
\Phi^\prime(0)= \langle \partial_t \Phi(0) \rangle \,.
\ee         
Finally,  the initial profiles for the local scale factor $e^\beta$ and
 expansion  rate $\beta^\prime$ are obtained from Eqs. (\ref{hubble2})
 and (\ref{momentum2}), with $\langle e^{\beta(0)} \rangle = 1$.  

\begin{figure}[t] 
\begin{tabular}{cc}
\hfil {\includegraphics[width=8cm,height=8cm]{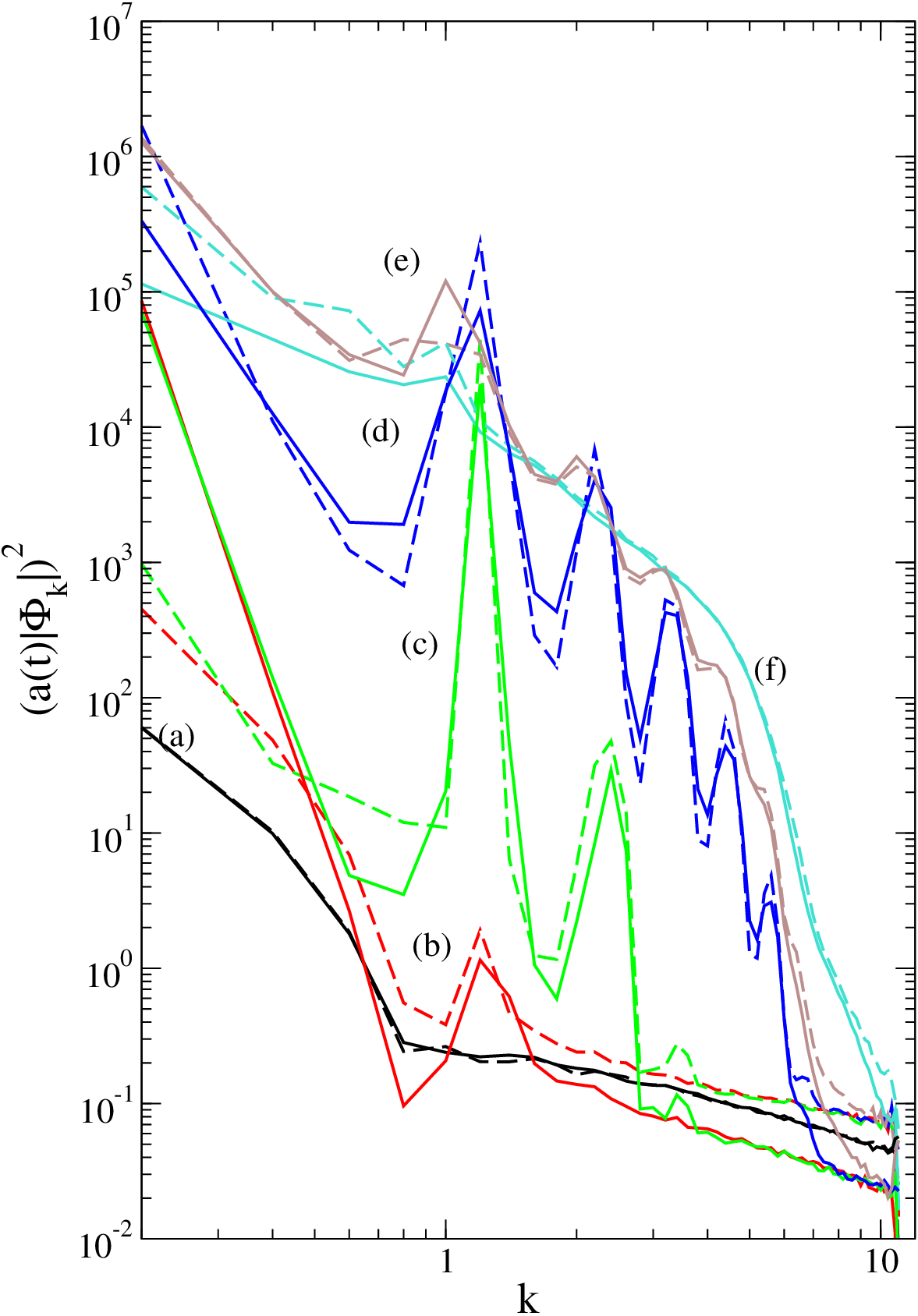}}\hfil
&
\hfil {\includegraphics[width=8cm,height=8cm]{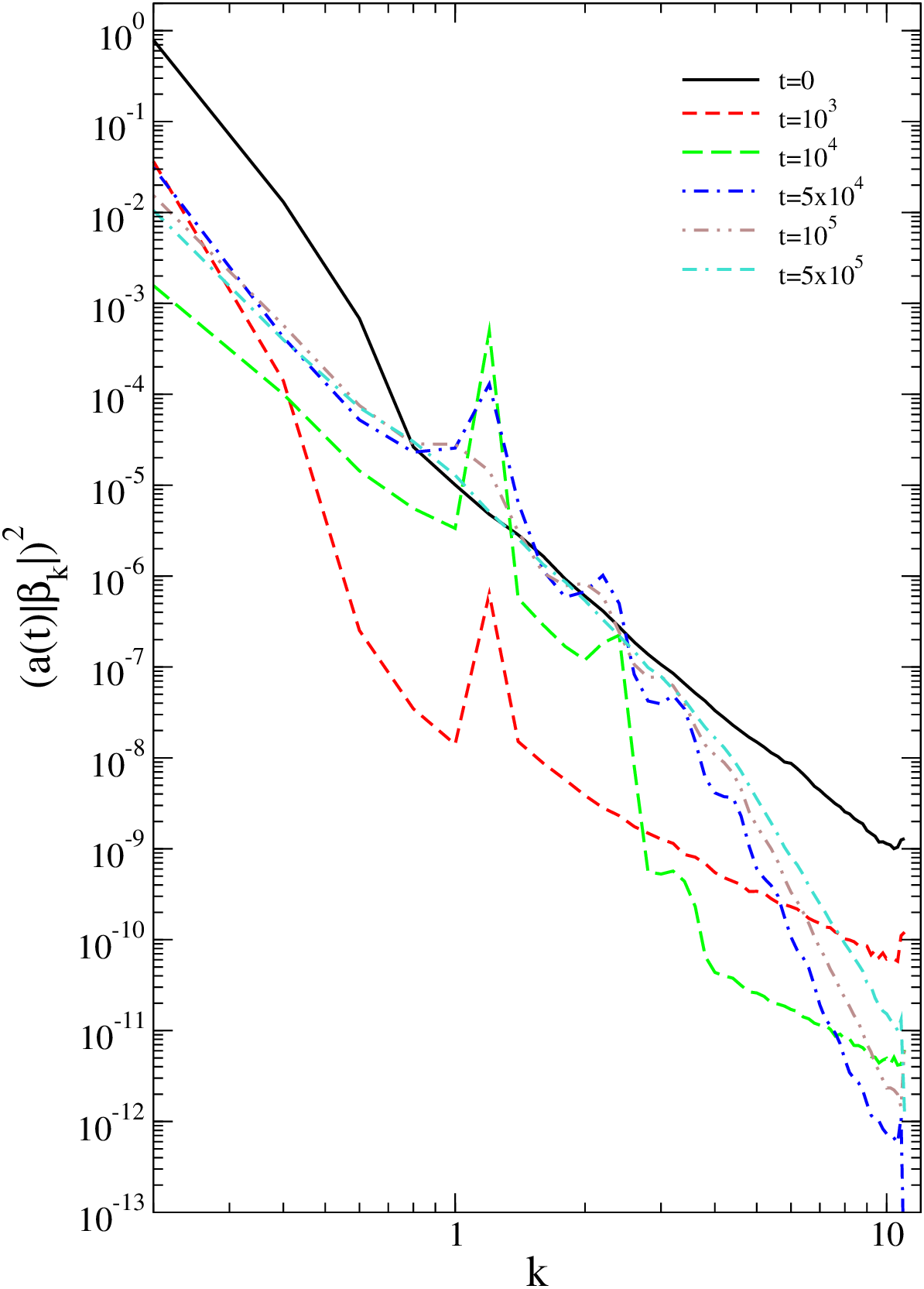}}\hfil
\end{tabular}
\caption{Left hand plot (a): Spectrum of inflaton field
  fluctuations, normalized by the initial value of the field, 
  $a(t)^2|\Phi_k|^2$, at different
  times, (a) $t=0$, (b) $t=10^3$, (c) $t=10^4$, (d) t=$5\times 10^4$,
  (e) $t=10^5$, (f) $t=5\times 10^5$. Solid lines are the spectrum with
  scalar metric fluctuations included. Dashed lines show the spectrum
  without metric fluctuations. Right hand plot (b): 
  Spectrum of scalar metric fluctuations $a(t)^2|\beta_k|^2$ at
  different times. We have defined $a(t)^2 \equiv \langle e^\beta \rangle^2$.  
}
\label{plot1}
\end{figure}

 We have run the simulations in a lattice with $N=64$ and $L=10
 \pi$, and periodic boundary conditions for the field. In program
 units, the smaller comoving wavenumber is $k=0.2$, 
 and the initial value of the average expansion rate  $H_0 = \langle
 \beta^\prime(0) \rangle \simeq 1/\sqrt{2}$. That means that at the
 beginning of the simulation there are a few field modes that have
 crossed the horizon before inflation ends, with $k < a_0 H_0$, with
 $a_0= \langle e^{\beta(0)} \rangle$, and have not yet reentered. For
 those modes the amplitude of the  spectrum has frozen around the time 
 $k = a_* H_*$, and therefore the initial value would be slightly
 larger than Eq. (\ref{initspectrum}) by a factor $a_0/a_*$
 \cite{moduli}, given by 
\be
\ln \frac{a_0}{a_*} \simeq \frac{3}{2} \left(\frac{a_0}{a_*}\,k-1
 \right) \,,
\ee
with $k$ given in program units. In Fig. (\ref{plot1}.a) is shown the
 spectrum of field fluctuations $|a(t)\Phi_k|^2$ at different times,
 where $a(t)=\langle e^\beta\rangle$ is the averaged scale factor. We
 compare the results obtained when including scalar metric
 fluctuations (solid lines) with those obtained by evolving the field
 with a background metric (dashed lines). At low values of the
 comoving wavenumber $k$ it can be seen the initial enhancement of the
 amplitude for the superhorizon modes, which is further amplified by
 the parametric resonance. But the main features of the resonance in
 this model are unchanged by the inclusion of superhorizon modes and/or
 metric fluctuations, and have been well established in the literature
 \cite{prechaotic4, lattice4-1,lattice4-2,turbulence}. First, field
 fluctuations grow in a narrow resonance regime, with the resonance
 peak located at $k\simeq 1.27$. At later times, rescattering effects
 lead to the appearance of multiple peaks, until finally the spectrum
 becomes smooth at small momentum, with an ultraviolet cut-off
 increasing in time.  It would be followed by a long stage of
 thermalization, during which the transfer of power from low to high
 momentum continues in the free turbulence regime \cite{turbulence}. 
 However, this stage is difficult to study numerically because of the
 limitations of the lattice cut-off, fixed by the choice of the
 lattice  size, which soon becomes smaller than
 the physical cut-off which is increasing with the scale factor. 

 The main difference when including metric fluctuations can be seen at
 smaller momentum during the first stages of the
 resonance. The initial superhorizon modes tend to be amplified at
 the same time than the main resonance peak, and this effect is
 enhanced by the metric fluctuations. This was already observed in
 Ref. \cite{precurvnum1} in the 1+1 dimensional system. Our
 simulations show that the effect prevails in 3+1
 dimensions. Nevertheless, at later times the tendency seems to be
 reversed, and the amplitude of the long wavelengths modes starts to be
 smaller than when integrating the system with a background
 metric. This might be an indication that due to the initial increase,
 the transfer of power from long to short wavelengths starts before
 and it reaches sooner the stage of free turbulence. 

 The spectrum of the scalar metric fluctuation $|\beta_k|^2$ is shown
 in Fig. (\ref{plot1}.b). The initial spectrum (top solid line) behaves like
\be
|\beta_k(0)|^2 \approx \kappa \frac{\omega_k}{k^4} |\Phi_k|^2 \,,
\ee 
for $k > 1$, showing the  same enhancement at low $k$ (superhorizon)
 than  that of 
 the field. With time, the spectrum is redshifted like the 
 inverse squared of the average scale factor $a(t)$, but it retains
 the resonance peak structure of the 
 field. The final spectrum (dot-dash-dashed line) is  given by the initial
 spectrum redshifted by a factor $a(t)^{-2}$, but with the initial power
 at low $k$ smoothed out, and showing the same cut-off at large
 $k$. The field amplitude in this model is redshifted like
 $a(t)^{-1}$, but the rescaled field mode $a(t)\Phi_k$ 
 exponential increase due to parametric resonance. This is
 not the case for the metric fluctuation, where the fluctuation is not
 really parametrically amplified.   

\begin{figure}[t]
\begin{tabular}{lr} 
 {\includegraphics[width=8cm,height=7cm]{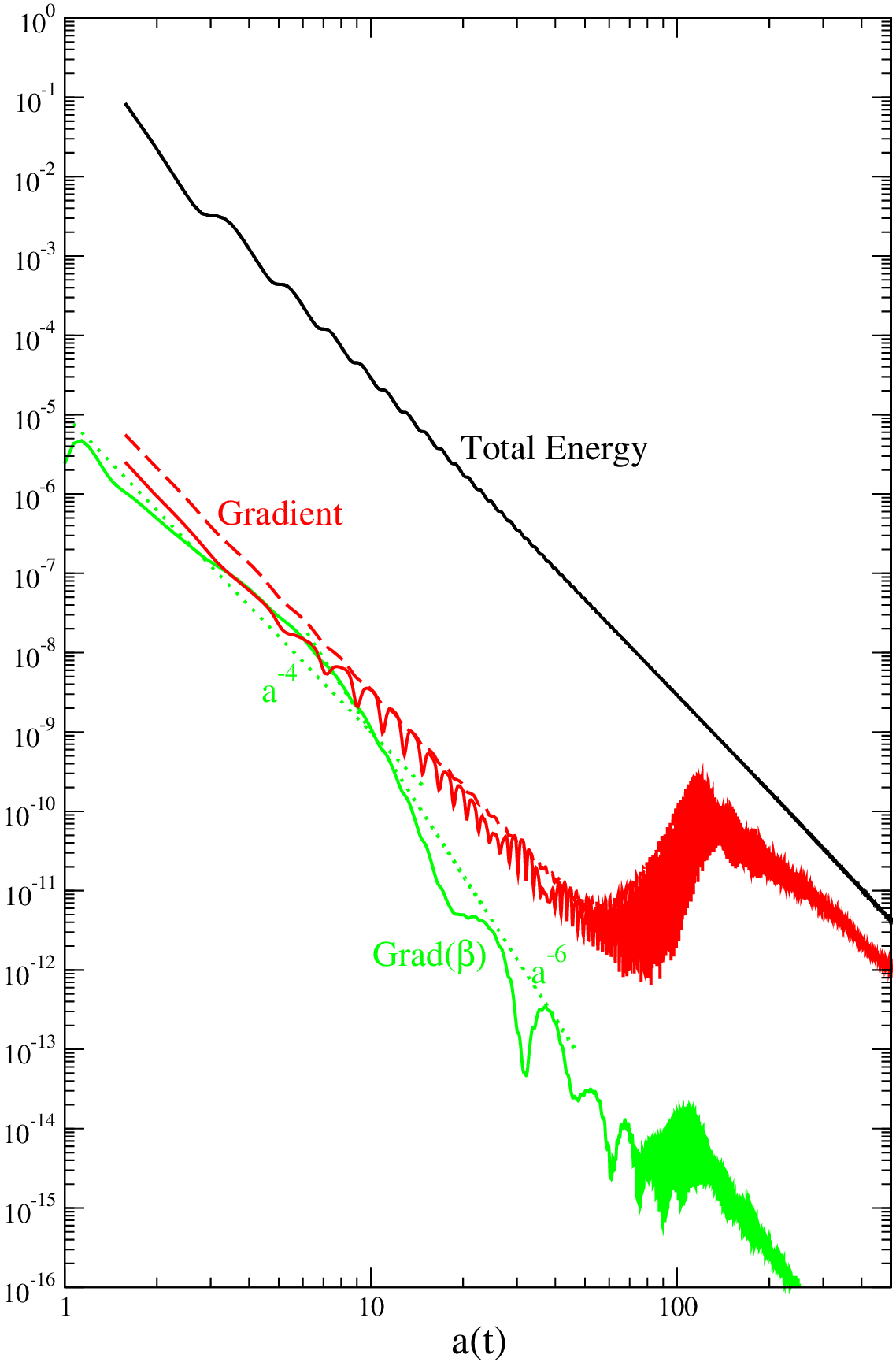}}\hspace{1cm}
&
\hfil {\includegraphics[width=8cm,height=7cm]{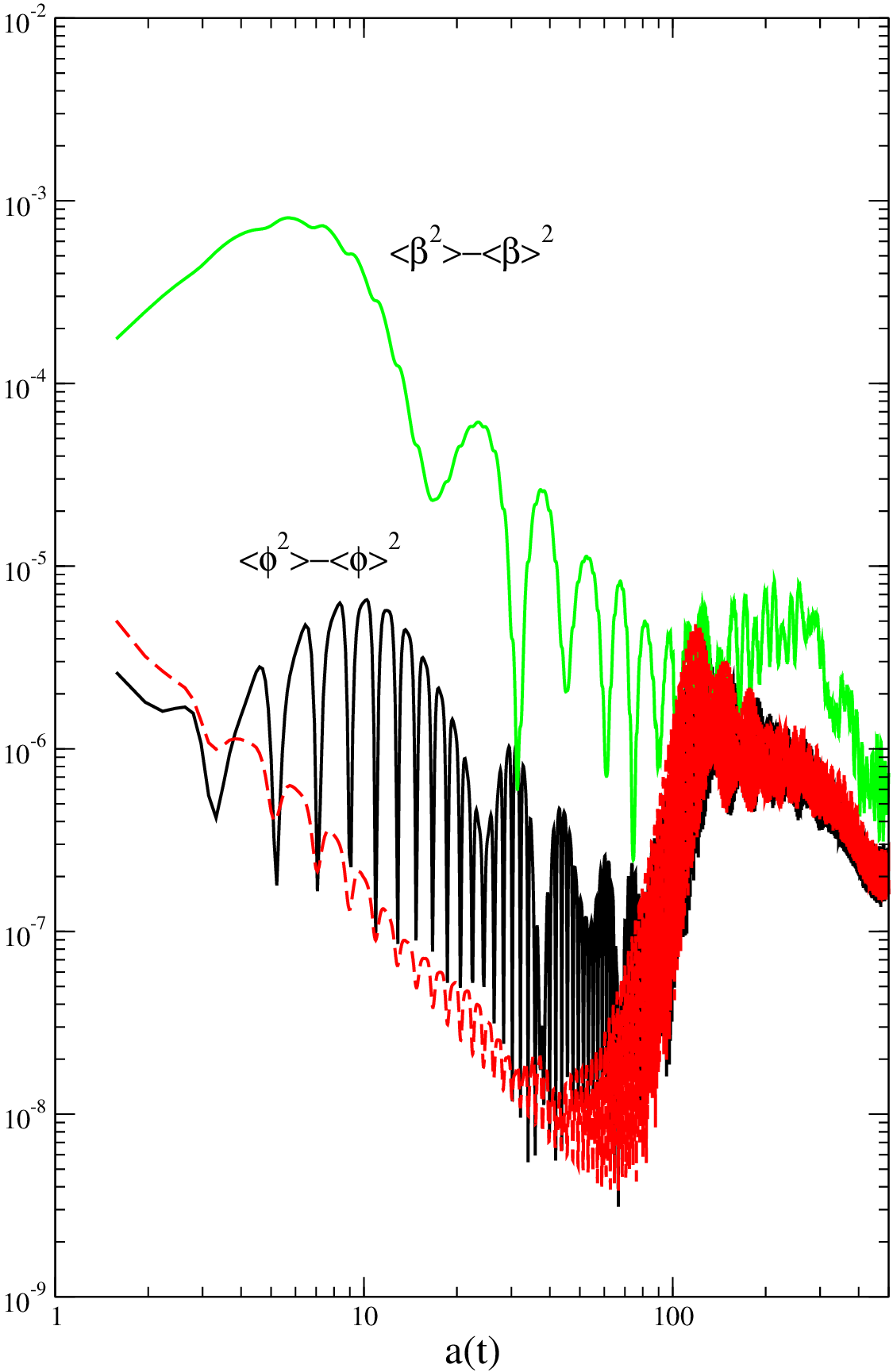}}
\end{tabular}
\caption{(a) Left hand plot: average total energy density $\langle
  \rho \rangle$ versus 
  the average scale factor $a(t)$ (black solid line). Also shown are 
  the average gradient contribution of the field Gradient=$\langle
  e^{-2 \beta} 
  |\nabla \Phi|^2/2\rangle$, and that of the scalar metric fluctuation
  Grad($\beta)=\langle e^{-2 \beta}|\nabla \beta|^2/2\rangle$. The
  dashed grey (red) line shows  
  the gradient of the field without metric perturbations. (b) Right
  hand plot: Variance of the field (solid bottom line) and scalar
  metric fluctuation (solid top line). The field variance without
  metric fluctuations (dashed line) is also shown for comparison.}  
\label{plot3}
\end{figure}

 When looking at the averaged values of fields, energies and variance,
 the inclusion of scalar metric fluctuations has a negligible
 effect. In Fig. (\ref{plot3}.a) is shown the average total energy
 density $\langle \rho \rangle$ with respect to the average scale
 factor. The system behaves as radiation with $\langle \rho \rangle
 \propto a(t)^{-4}$, and the average value is the same than that
 obtained without metric fluctuations. There is  at the
 beginning a slight difference in
 the gradient contribution of the field $\langle e^{-2 \beta} |\nabla
 \Phi|^2 \rangle$, but it practically disappears once
 the resonance broaden due to rescattering effects\footnote{For
 comparison with 
 the times at which the power spectrum is plotted, we have the
 standard time dependence of a radiation dominated universe, with
 $a(t) \sim \sqrt{t}$.} at around $a(t)\approx 50$. The gradient term of
 the metric fluctuation  $\langle e^{-2 \beta} |\nabla
 \beta|^2 \rangle$ initially follows that of the field, redshifted as
 $a^{-4}$, but the resonance in the field is not strong enough to
 keep this tendency and soon it starts decreasing faster as
 $a^{-6}$. Side by in Fig. (\ref{plot3}.b) we have included the
 variance of the field $\langle \Phi^2 \rangle - \langle \Phi
 \rangle^2$ (solid and dashed lines) and metric fluctuation  $\langle
 \beta^2 \rangle - \langle \beta 
 \rangle^2$ (solid line). Although the variance is initially
 larger with an inhomogeneous scale factor $\beta$, the difference
 again practically disappears when the resonance develops. The
 dispersion in $\beta$ is larger at the beginning than that of the
 field, but again it is less affected by rescattering effects. 

\begin{figure}[t]
 {\includegraphics[width=8cm,height=7cm]{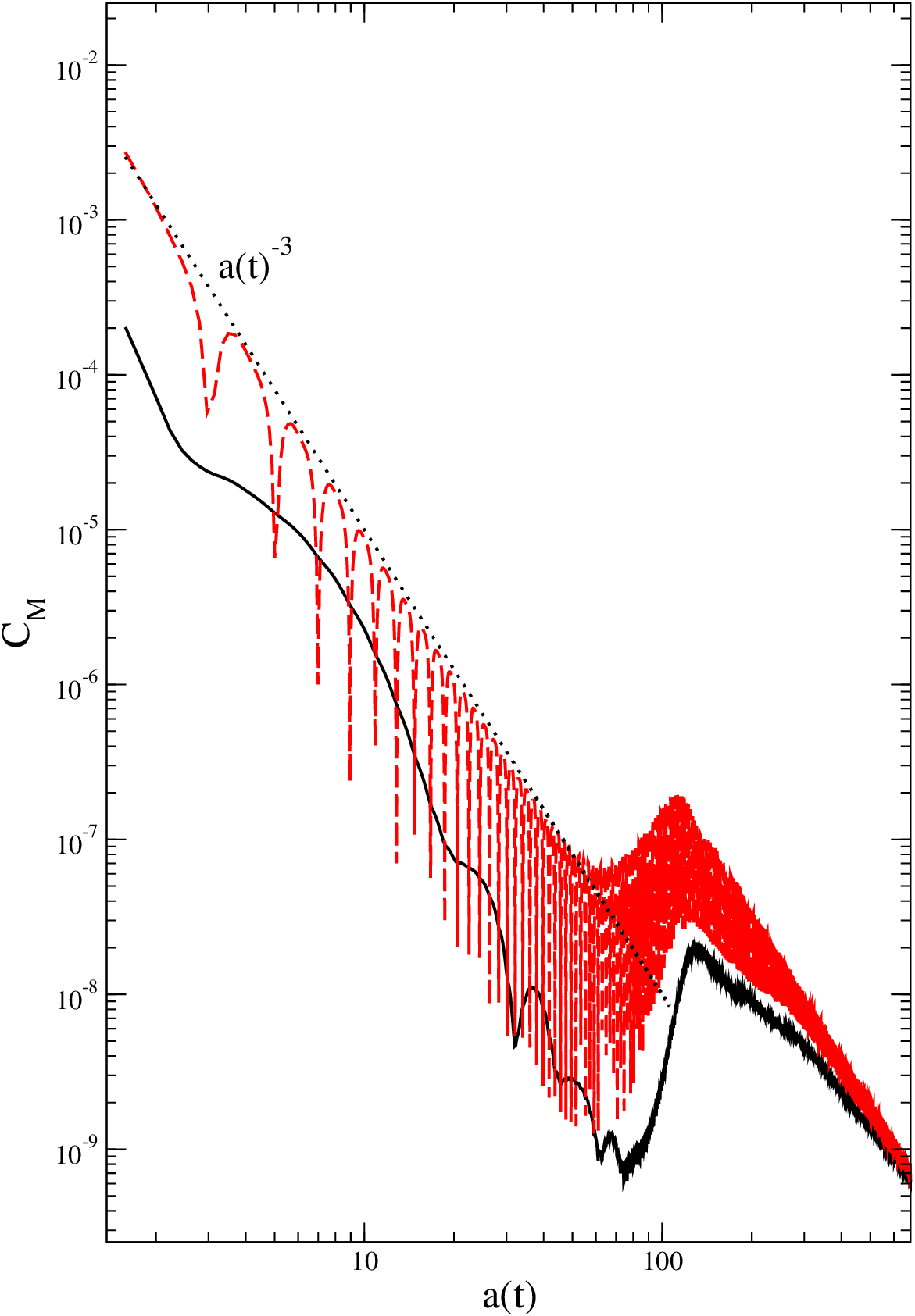}}\hfil
\caption{Average momentum constraint $C_M$ (see text), with (solid
  line) and without (dashed line) metric fluctuation. }  
\label{plot4}
\end{figure}

Finally, we have  checked to which extent metric vectors fluctuations
can be neglected in the field and scalar metric evolution, by checking
the momentum constraint Eq. (\ref{momentum2}). Although initially we
set the vectors and tensor fluctuations to zero, they are sourced by
scalar perturbations. The failure to satisfy this constraint during
the evolution can be interpreted as an indication of their
presence. This can be seen in Fig. (\ref{plot4}), where we have
plotted, as a function of the scale factor, $C_M$ defined by: 
\be
C_M= \langle |\partial_i \beta^\prime + \frac{\kappa}{2}\Phi^\prime
\partial_i \Phi|^2 \rangle^{1/2} \,, 
\ee  
which shows to which extend Eq. (\ref{momentum2}) is verified. This
will accounts for the terms neglected in Eq. (\ref{momentum}), such
that 
\be
C_M= \langle |\frac{1}{2}D_j A^j_i|^2 \rangle^{1/2} \,,
\ee
We include for comparison also the result without scalar metric
fluctuations, i.e., with an homogeneous scale factor. Although when
including $\beta$ the constraint $C_M$ seems to be better fulfilled by
and order of magnitude with less dispersion, in either case it just
follows the behavior of the field fluctuation, with $C_M$ decreasing initially
like $\langle \Phi^\prime \partial_i \Phi \rangle \propto a(t)^{-3}$
until the exponential increase due to the resonance of the
field. Indeed the behavior of $C_M$ indicates that  also the gradient
of the local Hubble parameter follows that of the field, with $\langle
\partial_i \beta^\prime \rangle \propto a(t)^{-3}$ instead of being
redshifted like $a(t)^{-2}$. We have checked that the behavior of
$C_M$ is not a numerical artifact due to discretization by integrating
the system for other choices of $L$ and $N$.   

Therefore, vectors (and by extension tensors) could be generated with an
initial amplitude similar to that of scalar fluctuations. However, due the 
behavior of the gradient term for $\beta$ in Fig. (\ref{plot3}.a), the source
term in their evolution equation (\ref{evoltraceless}) will be soon 
dominated by the contribution of the scalar field. Given
that the resonance is not strong enough to be fully transferred from the
fields to the scalar metric fluctuations, we might anticipate that the
same would happen for the other kinds of perturbations. Nevertheless, 
scalar metric fluctuations affect the field
spectrum at the start, although their effects are quickly redshifted, and
vectors and tensors may lead to a similar effect. It would be interesting
to study to which extent metric fluctuations apart from the scalar ones
help to enhance the power of the lower modes, whether this process
last for longer or just speed up the transfer of power through the
spectrum, but this is beyond the scope of the present work. 

\section{Hybrid model}
Preheating effects strongly depend on the model
considered, and models with an stronger resonance will lead to different
conclusion also about metric perturbations. Because of that, we study 
now  a different kind of model, a standard suspersymmetric hybrid model of
inflation, where it is known that the field resonance is stronger and
driven in what is called the tachyonic regime
\cite{prehybrid2,latthybrid,latthybrid-2}. The potential is given by:   
\be
V=V_0+ \frac{g^2}{4} N^4 + g^2(\Phi-\Phi_c^+)(\Phi-\Phi_c^-) N^2
+ \frac{1}{2}m_\phi^2 \Phi^2\,, \label{pothyb}
\ee
where as before $\Phi$ is the inflaton field, with a small mass
$m_\phi^2$, and $N$ is the waterfall field which triggers the phase
transition at  the end of inflation. Inflation takes place in the
false vacuum, with $\langle N\rangle \simeq 0$, and $V\simeq
V_0$. When the inflaton field overcomes 
the critical value $\Phi_c^\pm$, the squared field dependent
mass for $N$ becomes negative, due to the tachyonic instability
field fluctuations grow exponentially, and both fields move towards the
global minimum of the potential at $\phi_0=(\Phi_c^+ +\Phi_c^-)/2$ and
$N_0=(\Phi_c^+ - \Phi_c^-)/\sqrt{2}$. The potential in (\ref{pothyb})
reduces to the standard supersymmetric hybrid model when
$\Phi_c^-=-\Phi_c^+$, but it also includes other supersymmetric hybrid
potentials when the soft trilinear term is included
\cite{premar2}. Hereon we will take $\Phi_c^+= 2 \phi_0$, and
$\Phi_c^-=0$; other choices of these parameters will only shift the
background values of the fields at the global minimum, without
affecting the evolution of fields and field fluctuations.  
The vacuum energy $V_0$ driving inflation is adjusted such that the
potential energy vanished at the global minimum, i.e., $V_0 \simeq
g^2N_0^4/4= g^2 \phi_0^4$. As before, in order to eliminate the
coupling from the evolution equations we rescale the fields by $\phi_0$,
the potential and  energy density by $g^2 \phi_0^4$, time will be given in
units of $1/(g \phi_0)$, and wavenumber in units of $g \phi_0$. The
value of the coupling then sets the initial value of the fluctuations. 

Working only with the background fields during inflation, i.e. 
$\phi\equiv \langle \Phi \rangle$, the COBE amplitude of the primordial spectrum is given by:
\be
P_{\cal R}^{1/2} \simeq \left(\frac{H}{\dot \phi}\right)
\left(\frac{H}{2 \pi}\right) \simeq \frac{g}{4 \pi \sqrt{3} \eta_\phi}
\left(\frac{\phi_0}{m_P}\right)\,, \label{hybspec}
\ee
with $\eta_\phi = m_\phi^2/(3 H^2)$ one of the slow-roll
parameters, which also gives the tilt of the spectrum with the
spectral index $n_S \simeq 1 + 2 \eta_\phi$. Depending on
the values of the parameters in the potential, the coupling $g$ and
the mass $m_\phi^2$, the scale of hybrid inflation can range from the
unification scale, with $\sqrt{g} \phi_0 \simeq O(10^{15}$GeV), down to the
supersymmetric breaking scale $O(1$ GeV). Numerically, a too low value
of the scale means a lower value of the Hubble rate of expansion and
requires more integration time until the fields start 
oscillating. Therefore we will work here with models near the GUT
scale, and start with $\phi_0=0.005m_P\simeq 1.2 \times 10^{16}$ GeV,
$g=0.01$, $\eta_\phi=0.05$ and $V_0^{1/4}\simeq 1.2 \times
10^{15}$ GeV. With these values of the
parameters\footnote{Strictly speaking we also have a 
blue tilted spectrum with a rather large tilt, $n_S\simeq 1.1$,
excluded by observations \cite{WMAP,tensor-strings} when the tensor
contribution is negligible. We could always take a smaller value for
both the coupling and $\eta_\phi$ and satisfy all constraints. Our
choice of $\eta_\phi$ for this value of the inflationary scale is motivated by
numerical considerations, in order to shorten the first stage of the
integration from the end of inflation up to the point when
fluctuations starts growing.} we have $P_{\cal
  R}^{1/2} \simeq 5 \times 10^{-5}$, and a very small tensor-to-scalar
ratio $r \simeq 8 \times 10^{-6}$ . 

\begin{figure}[t]
\begin{tabular}{lr} 
 {\includegraphics[width=8cm,height=7cm]{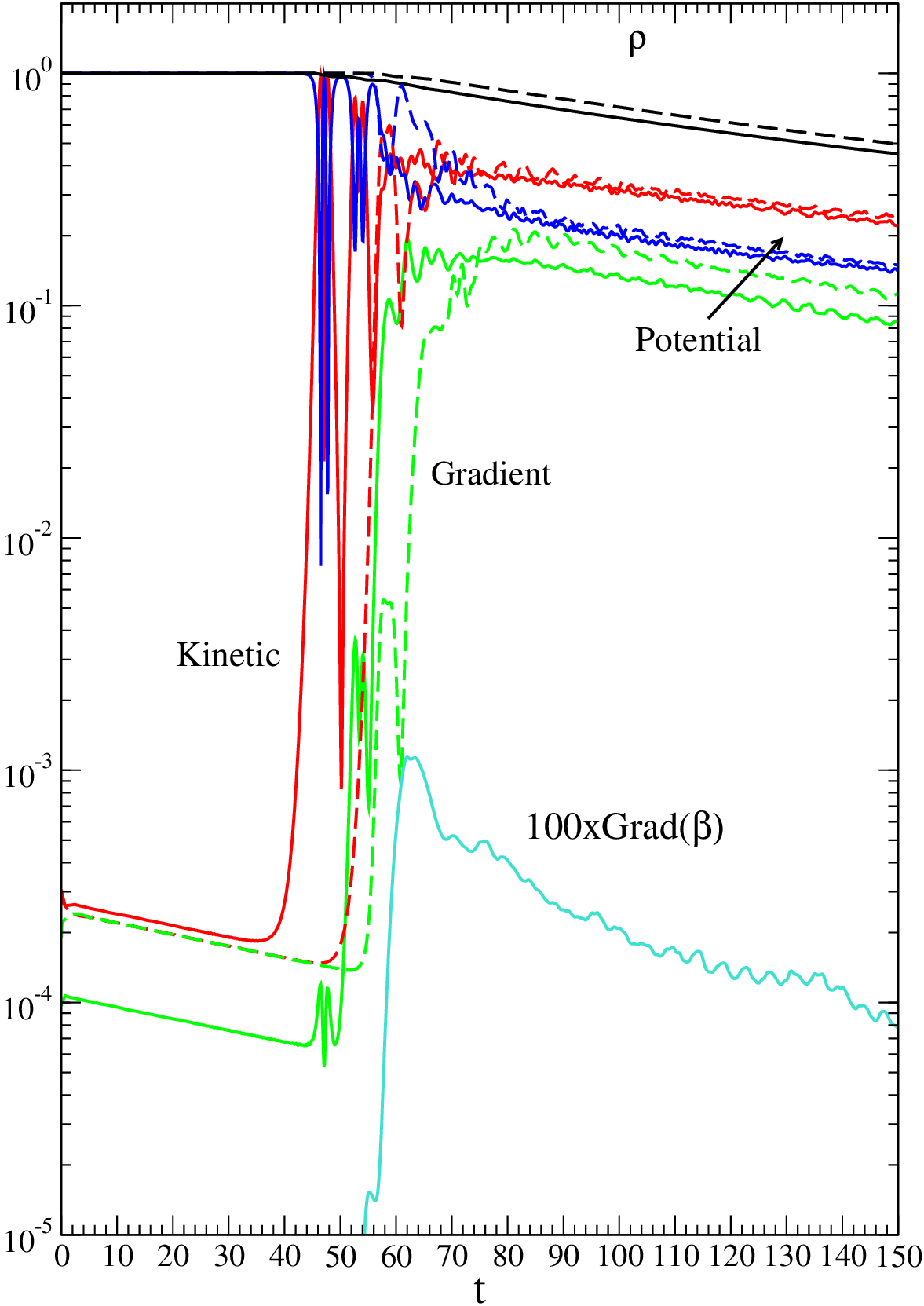}}\hspace{1cm}
&
\hfil {\includegraphics[width=8cm,height=7cm]{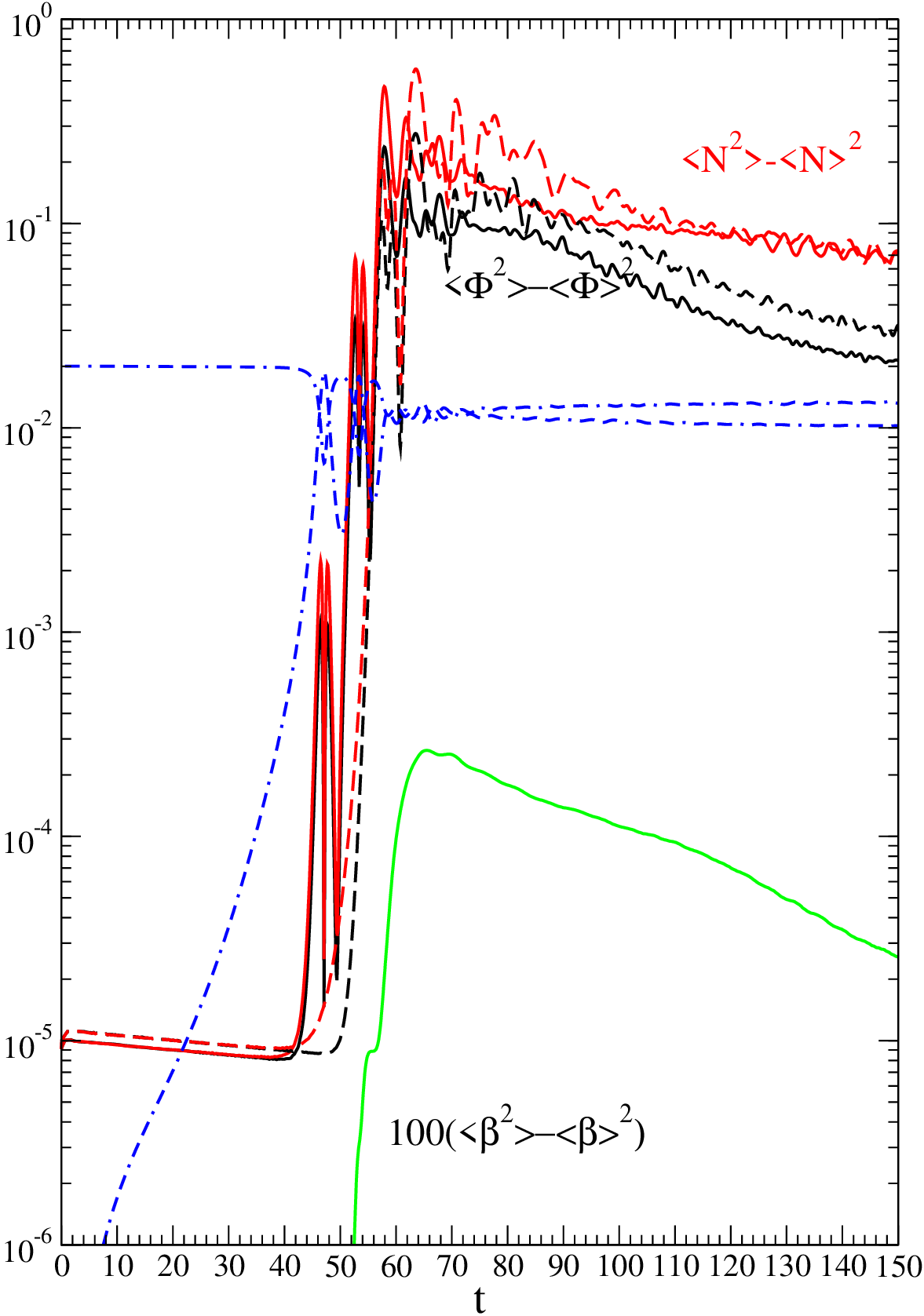}}
\end{tabular}
\caption{(a) Left hand plot: average total energy density  $\langle
  \rho \rangle$, potential $\langle V \rangle$, kinetic $\langle \dot
  \Phi^2 + \dot N^2 \rangle/2$, and gradient $\langle e^{-2
  \beta}((\nabla \Phi)^2 + \nabla N)^2)) \rangle/2$ energy density,
  versus time (solid   lines). The average gradient contribution of
  the scalar metric 
  fluctuation $\langle   e^{-2 \beta}|\nabla \beta|^2\rangle/2$ (solid
  grey/green line), scale by a factor of 100, is also shown.  The  
  dashed lines show the contributions without metric
  perturbations, total energy density, potential and field gradient
  energy density. Field
  quantities are given in units of the initial energy 
  density $g^2 \phi_0^4$, that of the metric fluctuation in units of
  $g^2 \phi_0^4/m_P^2$. (b) Right
  hand plot: Variance of the fields (solid lines), and scalar
  metric fluctuation scaled by a factor of 100 (solid grey/green bottom
  line). The field variance without
  metric fluctuations (dashed lines) is also shown for comparison. The
  dot-dashed lines are the  average field oscillations scaled by a
  factor of $10^{-2}$. Field quantities are given in units of  $\phi_0^2$, that
  of the metric variable in units of $(\phi_0/m_P)^2$. We have taken:
  $\phi_0=0.005\,m_P$, $\kappa=0.01$, $\eta_\phi=0.05$.    
}  
\label{plot5}
\end{figure}
We begin the evolution of the system just before the end of inflation,
i.e., some fraction of e-fold $N_e\simeq 0.05$ before the end with
the value of the background inflaton field just above the
critical value $\langle \Phi(0) \rangle = \Phi_c^+ e^{\eta_\phi
  N_e}$. We use the slow-roll condition to set the initial
value of the background velocity given by $\langle \dot \Phi(0) \rangle \simeq
-2 \eta_\phi \phi_0/m_P$ in program units. The background value of the
waterfall field $\langle N\rangle$ would be around zero, although
fluctuations very quickly drives this to a non zero
value. Numerically, it does not make much difference 
if we take a non vanishing initial value, and we have set $\langle
N(0) \rangle \simeq 10^{-8}$ and  $\langle \dot N(0) \rangle =0$. The
initial values of the inflaton fluctuations are given by 
Eqs. (\ref{phik}), (\ref{phidotk}) and  (\ref{initspectrum}). But now,  
 in order to fulfill the momentum constraint at $t=0$, we adjust the
 initial value of the $N$ fluctuations, identifying this field as the
 extra field considered in \cite{extrafield}. Therefore, we have taken:
\bea
N(0) &=& \Phi(0) - \langle \Phi(0) \rangle + \langle N(0) \rangle  \, \\
\dot N(0) &=& \langle \dot \Phi(0) \rangle - \dot \Phi(0)  \,
\eea
and the momentum constraint at $t=0$ is given by: 
\be
2 \partial_i \beta^\prime = -\kappa \langle \dot \Phi(0) \rangle
\partial_i \Phi(0) \,.
\ee         
With the choice of parameters given above, the Hubble rate of
expansion is rather small and practically the expansion effects are
going to be negligible during the tachyonic resonance. In particular, in
program units we have $H(0) \equiv \langle \beta^\prime(0) \rangle=
\phi_0/(\sqrt{3}m_P) \simeq 3 \times 10^{-3}$. This also means that at
the beginning of the evolution all modes considered are well inside
the Hubble radius. 

\begin{figure}[t]
\begin{tabular}{lr} 
 {\includegraphics[width=8cm,height=7cm]{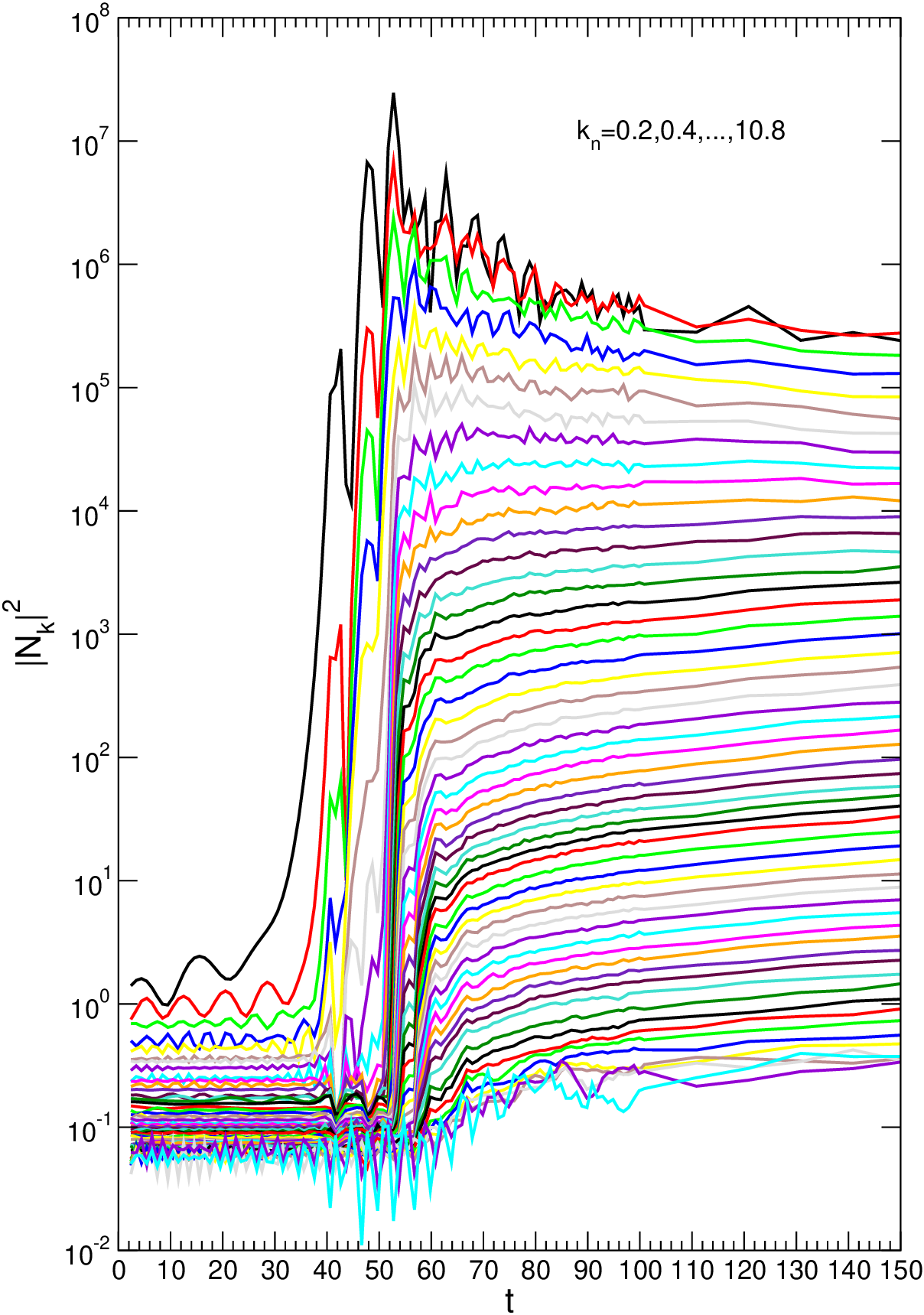}}\hspace{1cm}
&
\hfil {\includegraphics[width=8cm,height=7cm]{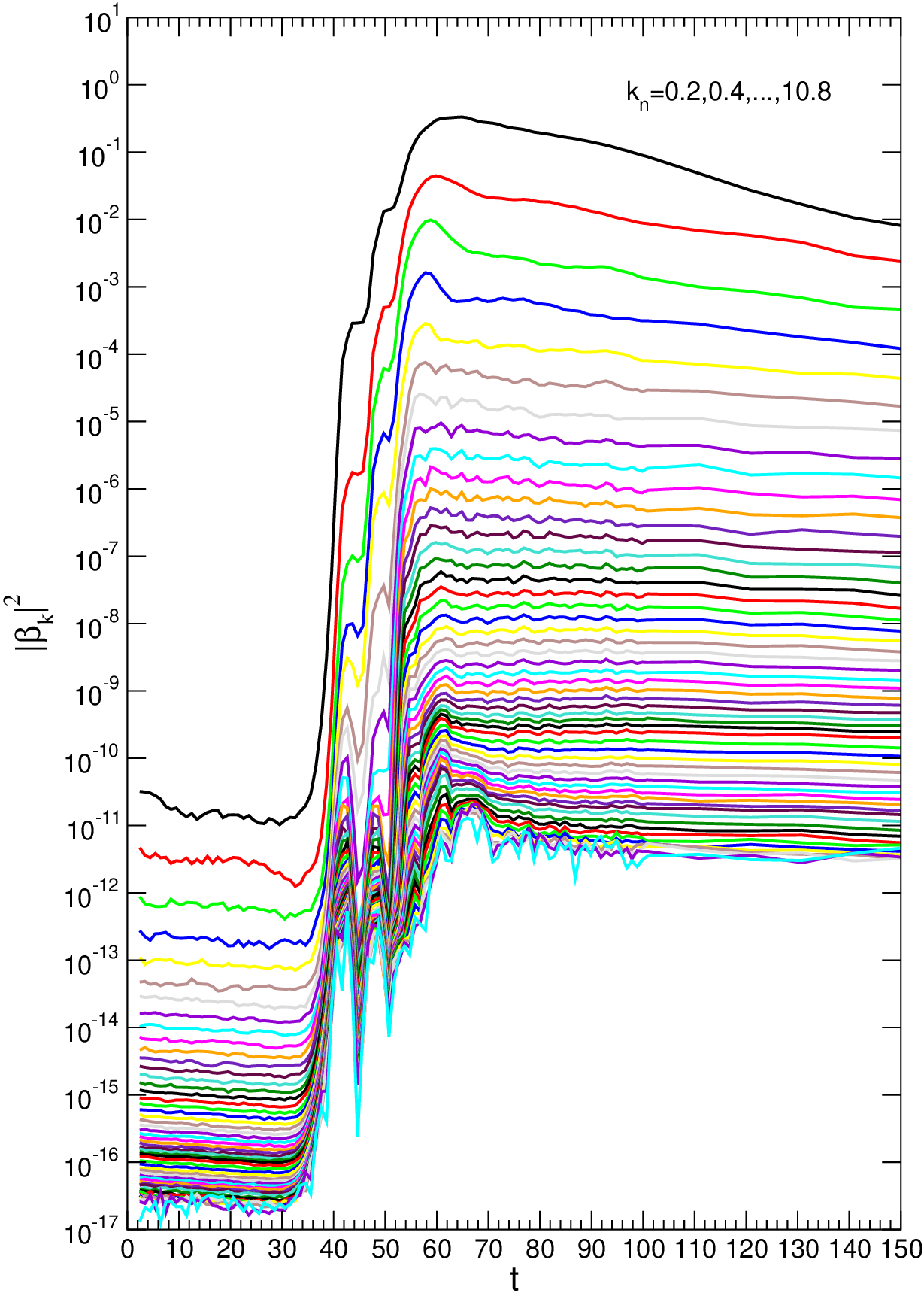}}
\end{tabular}
\caption{(a) Left hand plot: Time variation of the waterfall field
  spectrum $|N_k|^2$, in program units. (b) Right hand plot: Same for
  the scalar metric fluctuation  $|\beta_k|^2$.  We have taken:
  $\phi_0=0.005m_P$, $\kappa=0.01$, $\eta_\phi=0.05$.    
}  
\label{plot6}
\end{figure}

As before, we integrate the system in a cubic lattice with
$N^3=64^3$ sites and $L=10 \pi$, and periodic boundary conditions for
the fields. Preheating in this model proceeds
through the exponential growth of 
the waterfall field fluctuations due to the tachyonic instability in
the potential, which makes those of the inflaton  to grow at the
same rate \cite{prehybrid2,latthybrid}, and those of any other field coupled to
it \cite{premar2,latthybbaryon}. This lasts only a few oscillations of
the fields, depending on parameter values, during which the energy
density ends more or less equally distributed among potential, kinetic
and the gradient contribution of the fields. The tachyonic instability
ends when the field fluctuations of $\Phi$ and $N$ render the
effective squared mass for $N$ always positive during the oscillation,
i. e, when the variances of the fields become $O(\phi_0)$. Following
this there would be a period 
of bubble collisions \cite{latthybrid-2}, and then a stage of
turbulence with the transfer of momentum towards the ultra-violet, and
finally the thermalization of the system. As with the previous chaotic
model, we concentrate here in the first two stages without really
following the system into the turbulence regime. These can be seen in
Figs. (\ref{plot5}), where in the LHS we have plotted the total energy
density, potential, kinetic and gradient, and in the RHS the variance of the
fields. All quantities are given in program units, i.e., energy
densities are normalized to the initial vacuum energy $V_0$, and the 
variances to the field value $\phi_0$. We include for
comparison the results with no metric perturbations. There is no
qualitative difference between both cases, except for a slight shift
in time. In the RHS plot we have also included the averaged gradient
term for the scalar metric perturbation $\langle
e^{-2\beta}|\nabla\beta|^2\rangle$, rescaled by a factor of $100$. This
term is always negligible with respect to the field contributions,
and therefore metric perturbations has little or no effect on the
evolution of the fields. However, in this case the metric fluctuations
are also exponentially amplified, following the same resonance pattern
than that of the fields. This can be seen in Fig. (\ref{plot6}). On
the LHS it is shown the evolution with time of the waterfall field
spectrum $|N_k|^2$, and in the RHS that of $|\beta_k|^2$. Both
spectra behave during the resonance as $e^{2 \mu_k \Delta t}$, with a
growth index $\mu_k \sim 0.3$ for the lowest modes, slightly larger
for $\beta_k$. In addition, the size of our lattice $L=10 \pi$ is 
small enough to keep the physical ultraviolet cut-off in momentum for
the fields, with the largest momentum mode kept outside the resonance
band and being hardly enhanced.  
But this is not the case for the scalar metric fluctuations, for which all
modes are preheated.  
The initial scalar metric spectrum is
derived from the energy constraint Eq. (\ref{hubble2}), and therefore
$|\beta_k|^2 \propto O((\phi_0/m_P)^2)$. The exponential growth is not
enough, being only slightly larger than that of the field, to make the
scalar metric contribution in the evolution equations comparable to
that of the fields. In particular in the evolution equation for the
traceless metric components, Eq. (\ref{evoltraceless}), the source
term due to the scalar metric fluctuations would be of the order
$O(e^{-2\beta} |\nabla \beta|^2)$, and therefore suppressed by a
factor $O((\phi_0/m_P)^2)$ with respect to the field source
term.  For the same reason, the traceless components $A_i^j$ would be
of the order $\kappa e^{-2\beta} |\nabla \Phi|^2\sim
O((\phi_0/m_P)^2$, and their contribution to the evolution of fields
(\ref{evolfield}), and scalar metric fluctuation (\ref{evoltrace}), 
negligible.     

\begin{figure}[t]
 {\includegraphics[width=8cm,height=7cm]{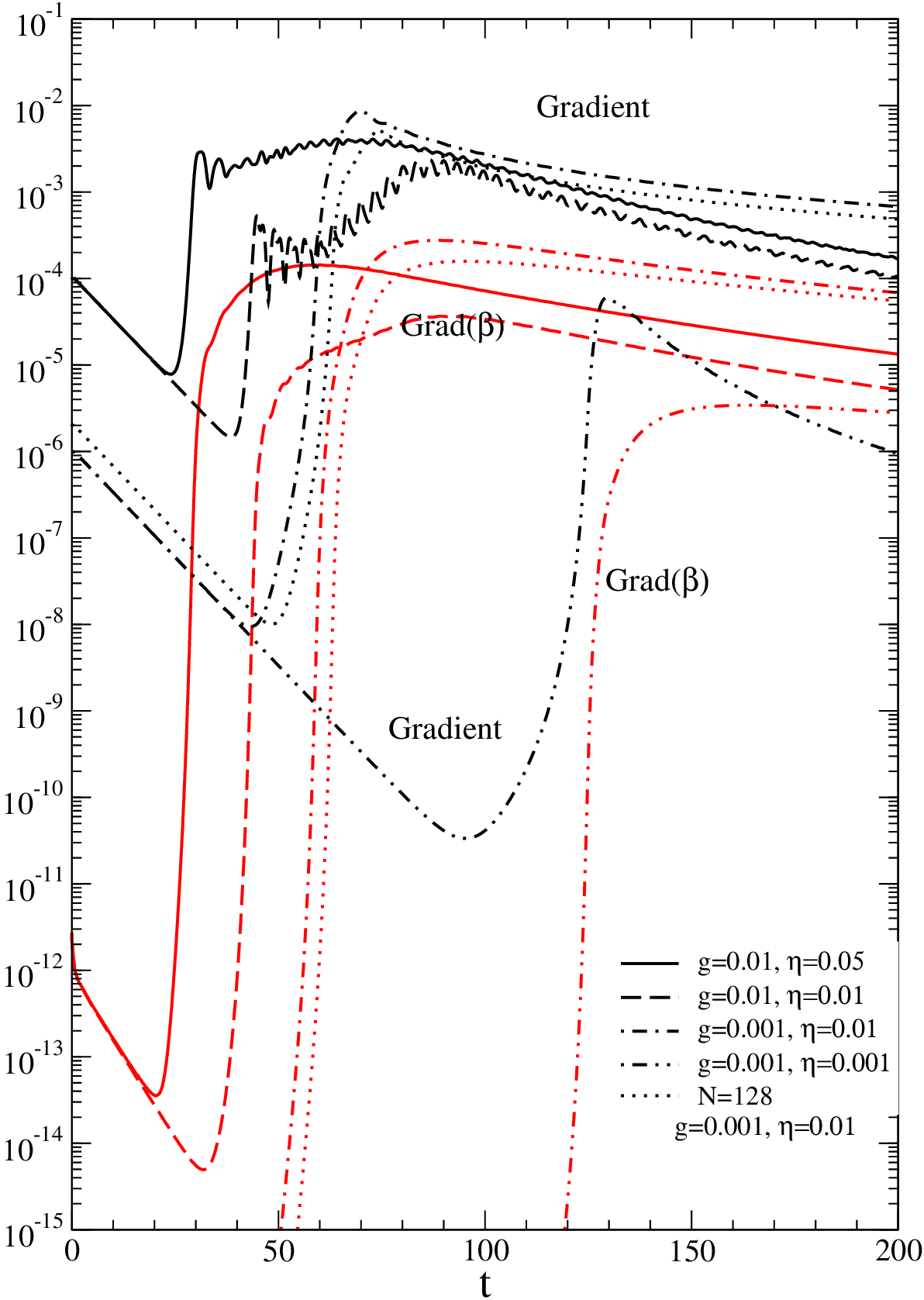}}\hfil
\caption{Comparison of the gradient of the fields (black lines),
  Gradient=$\langle e^{-2 \beta}((\nabla \Phi)^2 + \nabla N)^2))
  \rangle/2$, and that of the scalar metric fluctuation  (grey/red lines)
  Grad$(\beta)=\langle e^{-2\beta}(\nabla \beta)^2 \rangle/2$, 
  normalized by the initial energy density, for different parameter
  values and $N=64$, $L=10 \pi$:  
   $g=0.01$ and  $\eta_\phi=0.05$ (solid lines); $g=0.01$ and
  $\eta_\phi=0.01$ (dashed lines); $g=0.001$ and  $\eta_\phi=0.01$
  (dot-dashed lines); $g=0.01$ and  $\eta_\phi=0.001$ (dot-dot-dashed
  lines); in a lattice with
  $N=64$ and $L=10 \pi$. We also include for comparison the
  results obtained in a lattice with $N=128$, $L=10 \pi$, and 
  $g=0.001$, $\eta_\phi=0.01$  (dotted lines). For all we have taken
  $\phi_0=0.05m_p$. 
}  
\label{plot7}
\end{figure}

However, given that the main suppression of the scalar metric terms
with respect to those of the fields is due to factors of the order of 
$O(\phi_0/m_P)$, we have checked whether this scenario changes when 
increasing the rate of expansion, and/or decreasing the value of
$\eta_\phi$ and the coupling $g$. In particular, we want to check
whether or not the source term due to metric perturbations in the
evolution equation for the traceless modes Eq. (\ref{evoltraceless})
can be comparable to that of the fields. In Fig. (\ref{plot7}) we
have plotted the averaged value of the field's gradient 
and that of the scalar metric variable $\beta$, normalized by the
initial vacuum energy density, for different choices of the
parameters. In all of them we have increased the rate of expansion by
and order of magnitude, with $\phi_0/m_P=0.05$ and then $H_0\simeq 3
\times 10^{-2}$ in program units. Keeping first the other two parameters
the same than in the 
previous plots, $\eta_\phi= 0.05$ and $g=0.01$, we have already that
the scalar metric contribution becomes comparable to that of the
fields (solid lines). In this case, due to the larger Hubble expansion rate,
tachyonic preheating ends during the first oscillation of the fields.    
Decreasing the value of $\eta_\phi=0.01$, i.e., making the inflaton
potential flatter, slightly delays the beginning of the oscillations
and the start of the resonance (dashed lines), but qualitatively the
behavior of the gradient terms remain the same. Same when decreasing
the value of the coupling, with $g=0.001$ and $\eta_\phi=0.01$
(dot-dashed lines). By lowering  the value of the coupling we
decrease  the initial value (relative to the total energy density) of
the fluctuations. Until the fields start oscillating, they behave
approximately like massless fields, and the gradient terms are
redshifted like $a(t)^{-4}$, with the effective scale factor $a(t)
\equiv \langle e^\beta \rangle$. When further decreasing $\eta_\phi$
(dot-dot-dashed lines), the initial
value for the gradient terms is further decreased, but the relative
enhancement during the resonance remains the same. However in this case
the gradient of the fields decreases faster after the tachyonic
resonance  than that of $\beta$, such that the latter soon 
dominates after the resonance. 

Nevertheless, when increasing the scale of inflation we loose the physical
cut-off in our lattice, in the sense that the highest momentum mode is
further enhanced once tachyonic reheating ends. In this case, it just
take less than 
one oscillation for the variance of the fields to grow enough to
render the effective squared mass of the waterfall field positive. 
From the numerical simulations we have that for 
$g=0.01$ and $\eta_\phi=0.05$ this happens when $a(t)\simeq 2.3$; for
$g=0.01$ and $\eta_\phi=0.01$ when $a(t) \simeq 3.5$; and for
$g=0.001$ and $\eta_\phi=0.01$ when $a(t)\simeq 5.6$. Therefore,
practically all modes have been redshifted inside the resonance band
by the time the fields hit first the global minimum. In order to check
the dependence of the results on the lattice cut-off, we have run a
simulation with $N=128$
and $L=10 \pi$, i.e., doubling the maximum value of the comoving
momentum,  when $g=0.001$ and $\eta_\phi=0.01$, such that still at the
end of the resonance there are modes left outside the resonance
band. This is shown 
in Fig. (\ref{plot7}) by the dotted lines.  The values of
the gradients behave the same than before, and they do not differ by more
than a factor $O(2)$.  

\begin{figure}[h]
\begin{tabular}{lr} 
\includegraphics[width=8cm,height=7cm]{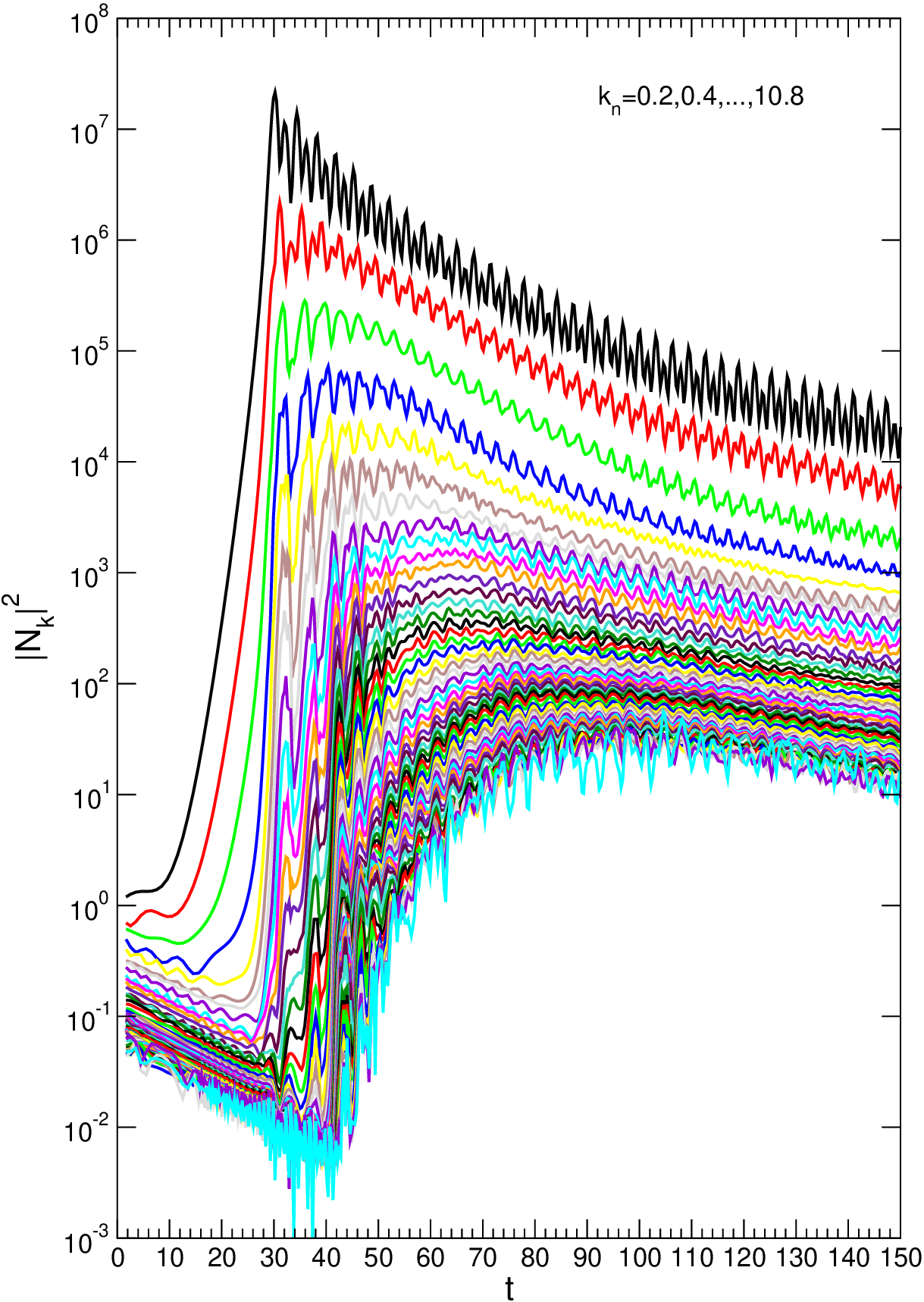}\hspace{1cm}
&
{\includegraphics[width=8cm,height=7cm]{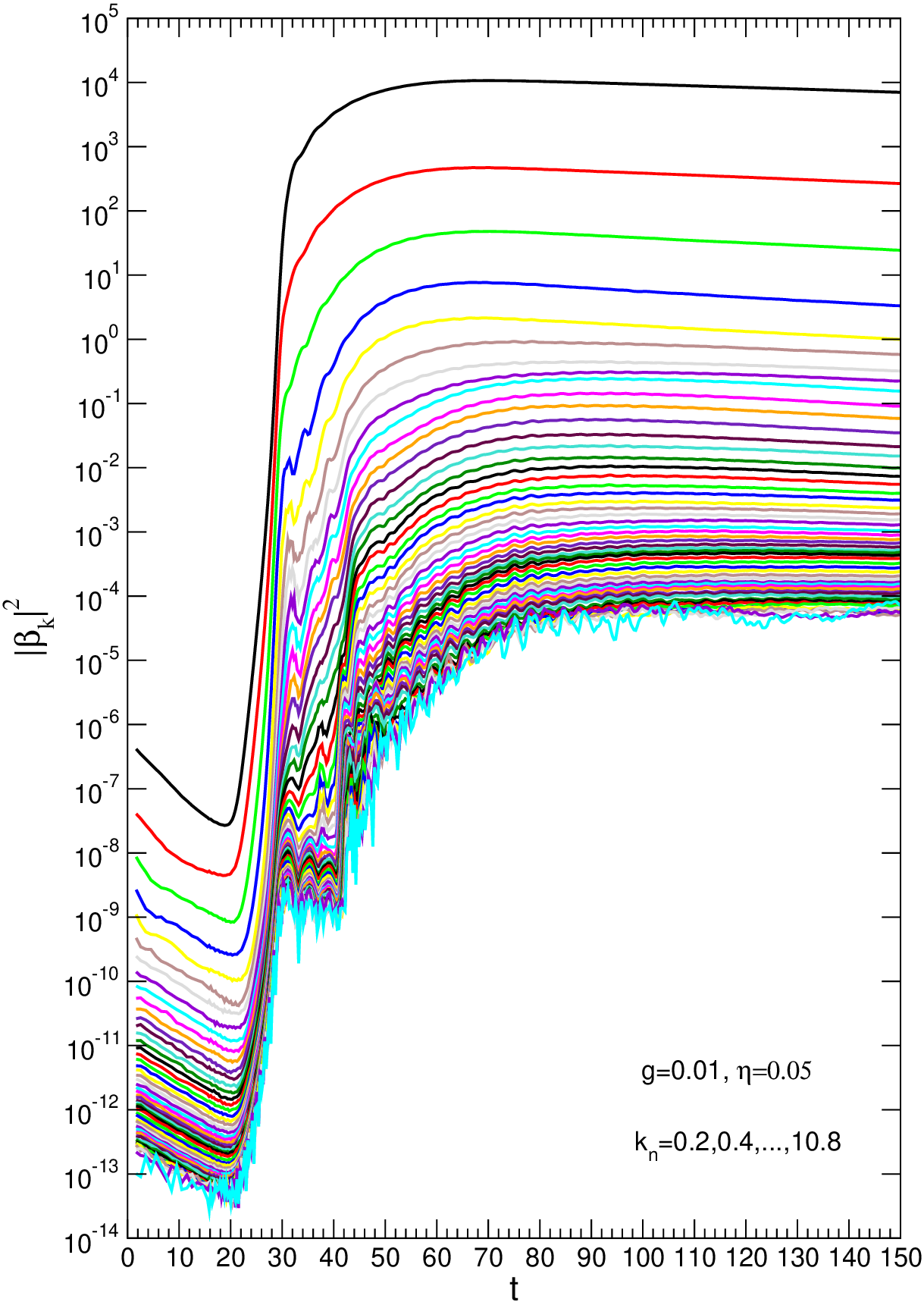}}
\end{tabular}
\caption{(a) Left hand plot: Time variation of the waterfall field
  spectrum $|N_k|^2$, in program units, for $\phi_0=0.05m_P$, $g=0.01$
  and $\eta_\phi=0.01$. $N=64$ and $L=10 \pi$ (b) Right hand plot: Same for
  the scalar metric fluctuation  $|\beta_k|^2$. 
}  
\label{plot8}
\end{figure}

In any case, during the short period of the resonance all long
wavelenghts modes are excited, and 
immediately after it seems to start a very effective transfer of
momentum from the infrared to the ultraviolet. This can be seen in
Fig. (\ref{plot8}), where as an example we have plotted the spectrum
of the waterfall field $|N_k|^2$ (LHS plot) and that of $|\beta_k|^2$
(RHS plot), for $\phi_0=0.05 m_p$, $g=0.01$, $\eta_\phi=0.01$. On the
latter, it can be seen more clearly the point at which the tachyonic
resonance ends at around $t\simeq 30$ when the highest modes stop to
be exponentially amplified, and the restart of the enhancement of the
highest modes at around $t \simeq 40$. Same happens for the other
model parameters considered in Fig. (\ref{plot6}), at different
time intervals. Given that the growth of the gradients practically ends with
the tachyonic resonance as expected,  we do not expect  their
evolution  to be much affected by the lack of high momentum modes. As
in the standard case without metric fluctuations, the resonance will
be followed by the turbulence regime, which study we do not pursue
here but it would be necessary in order to see the scaling followed by
the metric 
perturbation terms there. In Fig. (\ref{plot6}) the $\beta$ terms
decay as fast as that of the field, if not slower, but this tendency
remains to be further checked. But we would like to stress that, even
if non-linear metric perturbations are not preheated enough to affect
the evolution of the fields, they may become comparable to the field
source term in their own evolution equation, depending mainly on the
scale of inflation.  Our first study here indicates that this is the
case when $\phi_0/m_P > 0.05$, which sets the effective rate of
expansion during the numerical evolution, and it does not depend much
on the value of the coupling $g$, neither on that of the scale of
inflation.  For example a hybrid model with $\phi_0/m_P=0.05$ and
$\eta_\phi \simeq 0.01$, we will get the right amplitude of the
primordial spectrum with $g \simeq 10^{-4}$, which means a scale of
inflation $V_0^{1/4} \simeq 10^{15}$ GeV. The same scale is obtained with
$g=0.01$ and $\phi_0/m_P=0.005$, but in the former model we may
expect preheating of the metric perturbations to be non-negligible,
whereas in the latter we have checked that this is not the case.

\section{Summary}
\label{sect5}
We have studied the first stages of preheating following inflation in
a 3 dimensional spatial lattice, including both field and metric
perturbations. This extends previous one dimensional studies of the
problem as those of Refs. \cite{precurv1dim,precurvnum1,precurvrad}. 
 We have worked in the synchronous gauge, starting the evolution
of the system immediately after inflation. The set of evolution
equations for fields and metric variables are given by
Eqs. (\ref{evolfield}), (\ref{evoltrace}) and (\ref{evoltraceless}). 
During inflation, vector
modes decay, and tensor modes are subdominant with respect to the
scalar ones.  Thus we have taken them, vectors and tensors,  initially
to vanish. This means that we have to take initial profiles for the
field and field velocity consistent with the momentum constraint
Eq. (\ref{momentum}).  The non-linear evolution of the system couples
all kind of modes to each other, and any of them will be inevitably produced. 
Nevertheless, as a first step we have only followed the evolution of the
scalar metric perturbations and their effect on  the field
fluctuations. We have studied two different kind of models of
inflation, a chaotic model of inflation and a standard hybrid one,
which are representative of different patterns for the parametric
resonance. 
In the former, the resonance starts in the narrow resonance regime with
field mode fluctuations produced in a narrow range of comoving
wavenumber  \cite{preheating}. Later on, rescattering effects
re-distribute the 
resonance among several bands, for higher and lower comoving momentum
values \cite{lattice4-1,lattice4-2}. In the hybrid
model the resonance is driven by the tachyonic instability in the
waterfall field, and due to that it results in a more explosive and
faster production of particles \cite{prehybrid2, latthybrid,
  latthybrid-2}. In both cases, 
scalar metric perturbations hardly affect the evolution of field 
fluctuations, and the evolution of the field resonance is practically 
unchanged. For the chaotic model of Section II, however, there is an
initial enhancement of the field power at  lower wavenumbers, that
was also observed in Ref. \cite{precurvnum1}. In addition, we have also
included the effect of superhorizon modes in our simulations, with
larger initial amplitudes. Looking at the end of
the integration range in time, there might be also some
indication  that this initial transfer of power towards lower modes
helps in entering earlier the stage of free turbulence, where the
effect is reverse. But given the size of our numerical simulations we
cannot follow the system fully into that regime, and further studies
would be required in order to confirm this tendency. 

Although the field resonance does not feel the presence of metric
fluctuations, even in the non-linear regime, the scalar metric
perturbations follow the same resonance pattern than the field, as
seen in Fig. (\ref{plot1}.b) for the chaotic model, and
Fig. (\ref{plot6}.b) for the hybrid model. In the case of the chaotic
model the amplitude of metric perturbations is not really amplified
above the initial value, and the power spectrum of the metric
fluctuations simply reproduced the peak structure seen for the
field. This means that for the chaotic model metric fluctuations are
not really enhanced by 
the parametric resonance, they do not become comparable to the field
source terms in the evolution equations, and we can safely neglect
them. On the contrary, in the hybrid model the scalar metric
perturbations clearly feel the tachyonic resonance of the waterfall
field, in a similar way to any other field coupled to it like the
inflaton field. The scalar metric initial amplitude is amplified by
approximately the same factor $\sim e^{\mu_k \Delta t}$ than the
amplitude of the field fluctuation, with $\mu_k \sim 0.3$ being the growth
index. 
Although metric fluctuations never become large enough to
affect the field evolution, they may give rise to a source term 
comparable to that of the field in the evolution equation for the
traceless modes, Eq. (\ref{evoltraceless}), depending on the model
parameters $g$ and $\phi_0/m_P$, with $V_0^{1/4}= \sqrt{g}
\phi_0$. For example, this is the case for $g=10^{-4}$ and
$\phi_0/m_P=0.05$, i.e., for a scale of inflation $V_0^{1/4} \simeq 
10^{15}$ GeV. Whether this term helps to
enhance or not for example the production of gravitational waves
during preheating \cite{pregw,pregw2}, requires the study of the full
set of evolution equations for the metric perturbations, which is
beyond the scope of this paper and is left for further study.

\begin{acknowledgments}
We acknowledge the PLANCK collaboration for providing 
computing  resources. We have  used the SUNDIALS package, ``SUite of Nonlinear
and Differential, ALgebraic equation Solver'' \cite{sundials} for the
numerical integration.  
This work has been partially funded by  the MEC (Spain)- IN2P3 (France) 
agreement, ref. IN2P3 06-03.
.

\end{acknowledgments}

\end{document}